\documentclass[12pt]{article}
\usepackage{graphicx}
\usepackage[cp1251]{inputenc}  
\usepackage[russian]{babel}
\usepackage{latexsym}
\usepackage{amssymb}
\usepackage{amsmath}

\begin{document}

\makeatletter
\renewcommand*{\@cite}[2]{{#2}}
\renewcommand*{\@biblabel}[1]{#1.\hfill}
\makeatother

\title{New Interstellar Extinction Maps Based on Gaia and Other Sky Surveys}
\author{\bf \hspace{-1.3cm}\copyright\, 2023 г. \ \ 
G.A.Gontcharov$^1$\thanks{E-mail: georgegontcharov@yahoo.com},
\and \bf A.A.Marchuk$^{1,3}$,
\and \bf M.Yu.Khovrichev$^{1,3}$,
\and \bf A.V.Mosenkov$^{2,1}$,
\and \bf S.S.Savchenko$^{1,3,4}$,
\and \bf V.B.Il'in$^{1,3,5}$,
\and \bf D.M.Polyakov$^{1,3}$
\and \bf A.A.Smirnov$^{1,3}$
}
\date{$^1$ Pulkovo Astronomical Observatory, Russian Academy of Sciences, St. Petersburg, 196140 Russia \\
$^2$ Department of Physics and Astronomy, N283 ESC, Brigham Young University, Provo, UT 84602, USA \\
$^3$ St. Petersburg State University, St. Petersburg, 198504 Russia \\
$^4$ Special Astrophysical Observatory, Russian Academy of Sciences, Nizhnii Arkhyz, Karachai-Cherkessian Republic, 369167 Russia \\
$^5$ St. Petersburg State University of Aerospace Instrumentation, St. Petersburg, 190000 Russia}

\maketitle

\newpage

ABSTRACT
We present new three-dimensional (3D) interstellar extinction maps in the $V$ and Gaia $G$ filters within 2 kpc of the Sun, a 3D differential extinction 
(dust spatial distribution density) map along the lines of sight in the same space, a 3D map of variations in the ratio of the extinctions in the $V$ and Gaia $G$ filters 
within 800 pc of the Sun, and a 2D map of total Galactic extinction through the entire dust half-layer from the Sun to extragalactic space for Galactic latitudes $|b|>13^{\circ}$.
The 3D maps have a transverse resolution from 3.6 to 11.6 pc and a radial resolution of 50 pc. The 2D map has an angular resolution of 6.1 arcmin. We have produced these maps based 
on the Gaia DR3 parallaxes and Gaia, Pan-STARRS1, SkyMapper, 2MASS, and WISE photometry for nearly 100 million stars. We have paid special attention to the space within 200 pc of 
the Sun and high Galactic latitudes as regions where the extinction estimates have had a large relative uncertainty so far. Our maps estimate the extinction within the Galactic 
dust layer from the Sun to an extended object or through the entire dust half-layer from the Sun to extragalactic space with a precision $\sigma(A_\mathrm{V})=0.06$ mag. 
This gives a high relative precision of extinction estimates even at high Galactic latitudes, where, according to our estimates, the median total Galactic extinction through the
entire dust half-layer from the Sun to extragalactic objects is $A_\mathrm{V}=0.12\pm0.06$ mag. We have shown that the presented maps are among the best ones in data amount, 
space size, resolution, precision, and other properties. \\
DOI: 10.1134/S1063773723110026 \\
\bigskip\noindent
\leftline {PACS numbers: 98.35.Pr}
%
\bigskip

\newpage

\section*{INTRODUCTION}
\label{intro}

The spatial dust distribution, the corresponding stellar reddening, and the interstellar extinction are important for understanding the structure and evolution
of our Galaxy and extragalactic objects. The reddening of a star or the interstellar extinction between the observer and the star is determined most
accurately from its spectral energy distribution based on photometric, spectroscopic, and other observations. In this way, up-to-date data and methods
allow the individual reddenings/extinctions to be determined fairly accurately only for a minority of stars to which, for example, many non-single, peculiar, or
too dim stars do not belong. Furthermore, the error in the distance to a star introduces a considerable uncertainty into these results. As a result, the up-to-date
catalogues with reddening/extinction estimates for individual stars are very limited in both accuracy of these estimates and completeness of the sample of stars in any region of space.

For example, the data set by Anders et al. (2022, hereafter AKQ22)\footnote{https://data.aip.de/projects/starhorse2021.html or https://cdsarc.cds.unistra.fr/viz-bin/cat/I/354}
being used by us in this study was obtained from an analysis of the spectral energy distribution based on Gaia and other sky surveys and contains, among other things, individual 
extinction estimates for many stars. Within 2.5 kpc of the Sun the AKQ22 estimates were obtained for about 100 million stars from the total number of about 10 billion stars
in this space (Girardi et al. 2005). Naturally, the fraction of stars with measured individual reddenings/extinctions decreases with increasing distance from the Sun.

AKQ22 used the parallaxes and photometry from Gaia Early Data Release 3 (EDR3; Brown et al. 2021a) together with photometry from the Two Micron All-Sky Survey 
(2MASS; Skrutskie et al. 2006), the Wide-field Infrared Survey Explorer (WISE; Wright et al. 2010), the Panoramic Survey Telescope and Rapid Response System Data Release I 
(Pan-STARRS, PS1; Chambers et al. 2016), and the SkyMapper Southern Sky Survey DR2 (SMSS, SMSS DR2; Onken et al. 2019) to obtain comparatively accurate individual estimates 
of the distance $R$ and extinctions $A_\mathrm{V}$, $A_\mathrm{G}$, $A_\mathrm{BP}$ and $A_\mathrm{RP}$ in the $V$ and Gaia $G$, $BP$, and $RP$ filters, respectively.
The typical precision of the extinction estimates from AKQ22 is $\sigma(A_\mathrm{V})=0.15$ and 0.20 for bright and faint stars, respectively. It is approximately the same in
other individual estimates as well, for example, in the results of Berry et al. (2012). Given that the extinction itself near the Sun\footnote{Here and below, by the region near 
the Sun we mean the region with a radius of about 200 pc around it as a space where the typical uncertainty in the extinction is comparable to the extinction itself, 
while an insufficient number of stars does not allow many of the reddening/extinction determination methods, for example, the method of Green et al. (2019), to be applied.}
and at high latitudes is very low (for example, according to the recent estimates of the Gaia Collaboration (2023, hereafter TGE from the abbreviation of Total Galactic Extinction 
adopted by the Gaia Collaboration), the median total Galactic extinction through the entire dust half-layer toward the Galactic poles at $|b|>80^{\circ}$ is 
$A_\mathrm{V}\approx0.08\pm0.06$), it can be seen that the individual extinction of a star near the Sun and at high latitudes, as a rule, is known with an uncertainty greater than 100\%.

Therefore, despite the growth in the number of stars with individual reddening/extinction estimates, the estimates from reddening/extinction maps, which
can be more accurate than the individual estimates, remain topical.

A three-dimensional (3D) reddening/extinction map is a table that presents the reddening or extinction as a function of Galactic longitude $l$, latitude $b$, and distance $R$ 
from the Sun or rectangular Galactic coordinates $XYZ$.\footnote{We consider the Galactic rectangular coordinate system with the origin in the Sun and the $X$, $Y$, and $Z$ axes 
directed toward the Galactic center, in the direction of Galactic rotation, and toward the Galactic north pole, respectively. Examples of the cumulative and differential 
reddening/extinction maps as a function of $XYZ$ are the maps by Gontcharov (2017) and Lallement et al. (2022), respectively.}
As a rule, the estimates in the map are obtained by analyzing, averaging, and smoothing the individual estimates for stars in some space. The map-based estimate can be obtained
for any object by interpolating the estimates from the map cells adjacent to it. Thus, the map gives a reddening/extinction estimate for any object based on the estimates for 
its surrounding stars and the smoothing of natural dust medium fluctuations from star to star. Such fluctuations manifest themselves on a scale larger than 0.1 pc and have a typical
standard deviation from $\sigma(A_\mathrm{V})=0.06$ at high latitudes to $\sigma(A_\mathrm{V})=0.33$ near the Galactic equator, as discussed, for example, by Green et al. (2015),
Gontcharov (2019), Panopoulou et al. (2022), and Gontcharov et al. (2022).

Among the recent 3D maps based on Gaia but produced by different methods, we will note the maps by 
Gontcharov (2017, hereafter G17),\footnote{https://cdsarc.cds.unistra.fr/viz-bin/cat/J/PAZh/43/521} Green et al. (2019, hereafter GSZ19),\footnote{http://argonaut.skymaps.info/} 
Guo et al. (2021, hereafter GCY21), and Lallement et al. (2022, hereafter LVB22).

The 3D maps of differential reddening/extinction along the line of sight per unit distance are produced based on the 3D maps of cumulative reddening/extinction between the observer 
and a point of space. In fact, they represent the dust spatial distribution density variations.

In addition to 3D maps, 2D maps with estimates of the total Galactic reddening/extinction through the entire dust half-layer from the Sun to extragalactic space, i.e., 
as a function of only $l$ and $b$, are used. Such 2D maps are useful as a source of reddening/extinction estimates for extragalactic objects or for calibrating and testing the 
3D maps in which the cumulative reddening/extinction along the line of sight must be equal to the total Galactic reddening/extinction. It is the estimates from 2D maps that 
are most often used as Galactic reddening/extinction estimates in the most popular databases of extragalactic objects, such as the NASA/IPAC Extragalactic Database 
(NED; https://ned.ipac.caltech.edu). The 2D maps by Schlegel et al. (1998, hereafter SFD98) based on Cosmic Background Explorer (COBE) and Infrared Astronomical Satellite (IRAS) 
data and Schlafly and Finkbeiner (2011, hereafter SF11) based on the same data but with a different calibration (both are used in NED) as well as the map by 
Meisner and Finkbeiner (MF15; 2015) based on data from the Planck space observatory are most popular. These 2D maps are based on the estimates of the total dust emission in
the infrared (IR) on the entire line of sight followed by the elimination of the emission inside the Solar System and outside the Galaxy and the reddening--emission calibration.

In the current state of the art in astronomy the reddening/extinction maps are needed, first, as a source of estimates for numerous objects that do not have any estimates at all. 
For extended objects (star clusters, associations, interstellar clouds, galaxies, small regions of space, etc.) the map estimates are particularly accurate due to the smoothing 
of natural dust medium fluctuations. Note that hundreds of new open clusters and other extended objects, which are presented, for example, in the catalogues by 
Cantat-Gaudin et al. (2020) and Hunt and Reffert (2023), have been discovered in recent years. Second, the reddening/extinction maps are useful in that they help to detect and 
take into account the systematic errors of the individual estimates, as we show below using the distance dependence of the AKQ22 estimates as an example. Third, the 3D maps are 
needed to calibrate the observed quantities (for example, the IR emission) when producing the 2D maps (which, in turn, serve to test the new 3D maps).

Knowing the typical amplitude of the medium fluctuations, the typical error of an up-to-date map, and the typical accuracy of an individual extinction in present-day studies, 
one can estimate where the predictions of the maps for a point object are better and where the individual estimates for it are better. For example, the typical error of an 
individual extinction $\sigma(A_\mathrm{V})=0.18$ in AKQ22 is larger than the typical sum of the map error and the medium fluctuations in those regions where the fluctuations are
small or, more specifically, where they are $\sigma(A_\mathrm{V})<(0.18^2-0.08^2)^{0.5}=0.16$. This condition is fulfilled near the Galactic poles, approximately at $|b|>60^{\circ}$
(Gontcharov et al. 2022), where the map estimates are particularly useful. In the bulk of the sky, at $|b|<60^{\circ}$, the typical up-to-date individual extinction
estimates for point objects are more accurate than the predictions of up-to-date maps.

An overview and comparison of many 2D and 3D maps with an assessment of their advantages and disadvantages were provided, for example, by Gontcharov (2016b, 2017), 
Gontcharov and Mosenkov (2017a, 2017b, 2018, 2019, 2021a, 2021b), and Gontcharov et al. (2022). These studies show that the random and systematic errors of all up-to-date
2D and 3D maps are, at best, $\sigma(A_\mathrm{V})=0.08$, i.e., they are slightly lower than the typical errors of individual extinctions ($\sigma(A_\mathrm{V})=0.15$ for bright 
stars in AKQ22) but are still comparable to the extinction estimates themselves near the Sun and at high latitudes. Therefore, the discrepancies between the maps are also
particularly noticeable in these regions. For example, the estimates of the total Galactic extinction at $|b|>80^{\circ}$ from G17 and LVB22 are incompatible: $A_\mathrm{V}=0.18\pm0.07$
versus $0.02\pm0.03$, respectively (the random errors declared by the authors are given, while the actual errors, obviously, are much larger; see also Fig. 8 and its discussion).

The uncertainty in the emission–reddening calibration apparently makes a major contribution to the total uncertainty in the estimates from the most popular 2D maps 
(Gontcharov 2016b; Gontcharov and Mosenkov 2017b,2018, 2021a). The total uncertainty in the estimates from the 3D maps is formed from the set of uncertainties arising in the 
attempts to find a balance between the maximum resolution of the map and the maximum number of stars in each its cell. The uncertainty in the distances being used,
the incompleteness of the samples of stars under consideration, and the fallacy of the original assumptions about stars (for example, stars with a small reddening are erroneously 
considered as unreddened ones, binary stars as single ones, subgiants as dwarfs, etc.) are particularly influential.

So far no equally accurate and detailed 3D map both near and far (say, within a few kpc) from the Sun has been produced. Relatively accurate distances in the Gaia project can be 
obtained only for stars comparatively close to the Sun. For example, almost all of the Gaia stars have a comparatively high (better than 0.25) relative accuracy of their distances 
only within 2.5 kpc of the Sun. In contrast, expanding the space under consideration much farther than 2.5 kpc leads to the fact that stars with highly inaccurate distances prevail 
in the sample. They introduce large systematic errors into the estimates of the 3D map. Furthermore, an up-to-date 3D or 2D map based on optical photometry cannot extend too far 
from the Sun near the Galactic midplane, since the dust there obscures a considerable number of stars and distorts the estimates.

The estimates of the thickness of the Galactic dust layer obtained, for example, by Gontcharov and Mosenkov (2021b) show a noticeable growth of the extinction no farther than 
$|Z|\approx450$ pc from the Galactic midplane. Consequently, even when considering the extinction only within 2 kpc of the Sun, we nevertheless determine the total Galactic
extinction for all extragalactic objects at latitudes $|b|>\arcsin(450/2000)\approx13^{\circ}$.

All these considerations force us to pay special attention to the space near the Sun and at high latitudes far from the Galactic midplane when producing the 3D extinction map. 
For this purpose, we do the following: (i) construct maps within 2 kpc of the Sun using data for stars within 2.5 kpc of the Sun by taking into account the uncertainty in the 
distances; (ii) specify a coordinate grid step of 50 pc, i.e., uniform in distance rather than in distance modulus $D=5\log_{10}(R)-5$, as in many other maps covering a larger space 
but less detailed ones near the Sun; (iii) adopt angular and spatial resolutions of the map that provide an acceptable number of stars being used in all spatial cells; 
(iv) use dwarfs rather than giants in order that the sample be more complete near the Sun and the photometry be not overexposed.

In this study we present five maps produced by us based on AKQ22 data: 3D maps of extinction $A_\mathrm{V}$, extinction $A_\mathrm{G}$, and differential extinction $A_\mathrm{V}/R$
along the line of sight within 2 kpc of the Sun, a 3D map of variations in the extinction ratio $A_\mathrm{G}/A_\mathrm{V}$ that can be associated with the extinction law (i.e.,
the wavelength dependence of the extinction) within 800 pc of the Sun, and a 2D map of total Galactic extinction $A_\mathrm{V}$ through the entire dust half-layer from
the Sun to extragalactic space for latitudes $|b|>13^{\circ}$.

\begin{figure*}
\includegraphics{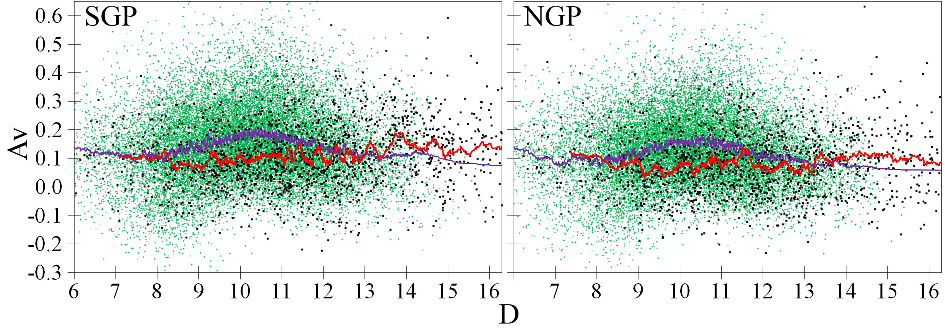}
\caption{Extinction $A_\mathrm{V}$ for the dwarfs (the green symbols and the violet curves of the moving average over 150 points) and giants (the black symbols and the red curves 
of the moving average over 40 points) from AKQ22 within 4 degrees of the south (SGP) and north (NGP) Galactic poles versus distance modulus $D$.
}
\label{sgpngp}
\end{figure*}

\begin{figure*}
\includegraphics{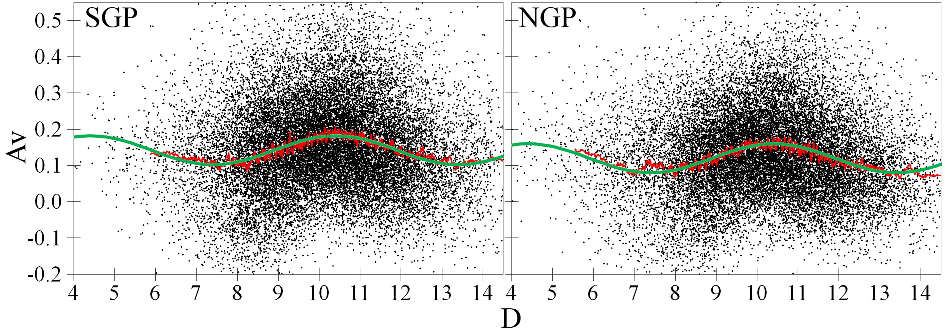}
\caption{Same as Fig. 1 but only for the dwarfs (the black symbols and the red curves of the moving average over 150 points).
The green curve indicates the fit by Eqs. (1) and (2) for the SGP and NGP, respectively.
}
\label{dwarfs}
\end{figure*}

\section*{DATA AND METHOD}
\label{data}

The estimates of the distance R and extinction $A_\mathrm{V}$ from AKQ22 served as the main data for our 3D and 2D maps. Let us briefly describe them.

As noted in the Introduction, the parallaxes and some other characteristics of stars from Gaia EDR3, photometry in the $G$, $BP$, and $RP$ bands from Gaia EDR3, 
$g_\mathrm{PS1}$, $r_\mathrm{PS1}$, $i_\mathrm{PS1}$, $z_\mathrm{PS1}$ and $y_\mathrm{PS1}$ from PS1, 
$g_\mathrm{SMSS}$, $r_\mathrm{SMSS}$, $i_\mathrm{SMSS}$ and $z_\mathrm{SMSS}$ from SMSS,
$J$ and $Ks$ from 2MASS, $W1$ and $W2$ from the AllWISE catalogue (one of the versions of the WISE results) for the same Gaia EDR3 stars, and the declared AKQ22 uncertainties of 
all the estimates used served as the input data for the estimates in AKQ22. These data were calibrated, corrected, and selected by AKQ22 in accordance with the recommendations of
the authors of the Gaia, PS1, SMSS, 2MASS, and AllWISE data sets used. In particular, the parallaxes were calibrated in accordance with the recommendations of the Gaia Collaboration 
(2021c), while the accurate photometry was selected in accordance with the criteria of the Gaia Collaboration (2021b). While processing the data, AKQ22 formulates their
several criteria for the selection of high-quality $R$ and $A_\mathrm{V}$ estimates. We used the AKQ22 estimates with the criteria \verb"fidelity" $>0.5$ (sufficiently accurate
stellar astrometry), \verb"s.sh_outflag='%%00'" (reasonable uncertainties in the quantities being used), \verb"(av84-av16)/2<0.25" (an accuracy of the extinction better than 0.25), 
and \verb"(dist84-dist16)/2/dist50<0.25" (a relative accuracy of $R$ better than 0.25). We selected the dwarfs according to the criterion \verb"logg50" $>3.95$ for the surface gravity.

To calculate $R$ and $A_\mathrm{V}$ from the input data, AKQ22 used the StarHorse code described by Queiroz et al. (2018). This code calculates the most probable estimates
of $R$, $A_\mathrm{V}$, age, mass, effective temperature, metallicity, and surface gravity for a star by comparing the theoretical PARSEC1.2S+COLIBRIS37 isochrones 
(Bressan et al. 2012) and the input measured quantities on the color–magnitude diagrams. Justified initial constraints are imposed on the quantities being determined in accordance 
with the views of the structure and evolution of stars in the Galaxy (for example, the initial mass function) and such components of the Galaxy as the thin and thick disks,
the spherical halo, and the triaxial bulge–bar structure. The isochrones being used were calculated only for the solar metallicity scale, without taking into account the enrichment 
with alpha-elements in the Galactic halo.

For the estimates being used by us it is important that AKQ22 imposed wide initial constraints on the $A_\mathrm{V}$ estimates in accordance with the reddening estimates
from the GSZ19 map and the 3D model of spatial reddening variations by Drimmel et al. (2003, hereafter DCL03). As noted by the authors of GSZ19 and DCL03 themselves, they give 
unsatisfactory estimates near the Sun. Furthermore, it was shown in the papers of Gontcharov and coauthors mentioned in the Introduction that the GSZ19 and DCL03 estimates are also 
unsatisfactory at high Galactic latitudes. Consequently, the GSZ19 and DCL03 errors can lead to errors in the AKQ22 estimates. One of the AKQ22 systematic errors is considered below.

Figure 1 shows the systematic extinction variations with distance modulus $D$ for the dwarfs and giants selected by us from AKQ22 within 4 degrees of the south (SGP) and north (NGP) 
Galactic poles. For both dwarfs and giants the extinction variations are not monotonic even after a strong smoothing. In addition, for both classes of stars the extinction changes
systematically even at $D>10$, i.e. at $R>1000$ pc, or $|Z|>1000$ pc, given that the neighborhoods of the poles are considered. This is totally inconsistent with the views of a 
Galactic dust layer thickness $|Z|<450$ pc (Gontcharov and Mosenkov 2021b). Consequently, the estimates for both classes of stars are distorted by systematic errors. 
The spatial density of giants within 2.5 kpc of the Sun is quite insufficient both for the analysis of these errors and for the construction of detailed and accurate 3D extinction
maps. Therefore, in this study we use only dwarfs or, more specifically, 99\,889\,339 dwarfs from AKQ22 within 2.5 kpc of the Sun. Note that 1\,171\,388 (1\%) of them have negative 
extinction estimates in AKQ22, reflecting the fact that the star can also be bluer than the corresponding isochrone on the color--magnitude diagram because of the stellar photometry, 
distance, classification errors and other factors.

Figure 2 copies Fig. 1 but only for the dwarfs. The systematic extinction variations are seen to be well fitted by the sine waves found by us:
\begin{equation}
\label{sgpsin}
A_\mathrm{V}=0.14+0.04\sin(1.0467D-3.035)
\end{equation}
\begin{equation}
\label{ngpsin}
A_\mathrm{V}=0.12+0.04\sin(1.0467D-3.035)
\end{equation}
for the SGP and NGP, respectively, where the argument of the sine is expressed in radians.

This systematics affects the results of any study using the AKQ22 data, especially near the Sun and at high latitudes, where the amplitude of this systematics is comparable to 
the extinction estimates themselves. Undoubtedly, this requires a separate study. We found no unequivocal explanation of this systematics, but we will note the following.

AKQ22 pointed out that the broad-/intermediate-band photometry used by them is only marginally sensitive to metallicity. This can lead to large errors in the metallicity of the 
stars for which AKQ22 found a low metallicity is shown in their Fig. 10, is discussed by AKQ22, and does not correspond completely to the universally accepted views of the Galaxy. 
Moreover, a comparison of the metallicity estimates from AKQ22 with the definitely more accurate astroseismic and spectroscopic estimates in Figs. 17 and 18 from AKQ22 and in their
discussion shows that AKQ22 could systematically underestimate the metallicity of a considerable number of stars. Given that low-metallicity stars are, on average, systematically 
bluer and fainter than high-metallicity ones, the systematics in Fig. 2 can be explained as follows. AKQ22 may erroneously consider many of the high-metallicity stars as low-metallicity
ones, and their observed comparatively red color is then erroneously interpreted as a large reddening of a low-metallicity star with a comparatively blue dereddened color instead 
of recognizing a small reddening of a high-metallicity star with a comparatively red dereddened color. In contrast, low-metallicity stars may be erroneously considered as 
high-metallicity ones. This can give not one but several extrema in the extinction variations with distance, because AKQ22 divide the stars into the thin disk, the thick disk, and
the halo. Another probable explanation of the mentioned systematics is the influence of numerous unresolved binary dwarfs that is ignored in AKQ22. Such dwarfs look brighter and 
redder than is suggested by the model of a single dwarf. Similarly to the error with the metallicity, their observed comparatively red color is erroneously interpreted as a 
large reddening of a single star with a comparatively blue dereddened color. In this case, the sinusoidal dependence of the error on distance is explained by the maximum
influence of unresolved binaries at some distance: at a smaller distance there are more resolved binaries, while at a larger distance the influence of unresolved
binaries must be smaller than the photometry errors.

As a result, we eliminated this systematics by adopting the empirical correction $\Delta A_\mathrm{V}=-0.04\sin(1.0467D-3.035)$. Thus, we take the mean values of the sine waves 
as the true ones and, accordingly, keep the mean extinction for the dwarfs unchanged. This seems reasonable, since at $D>11$ the mean extinctions for the dwarfs and giants are
close, as can be seen from Fig. 1. However, at $8<D<11$ the $A_\mathrm{V}$ estimates for the giants are, on average, lower than those for the dwarfs by 0.04 mag.
If it will emerge in future than these estimates for the giants are closer to the truth, then the $A_\mathrm{V}$ estimates in the maps presented by us need to be reduced by 0.04 mag.

The mean extinction through the entire dust half-layer toward the SGP or NGP from the AKQ22 dwarfs is $A_\mathrm{V}\approx0.14$ and 0.12, respectively. The difference 
$\Delta A_\mathrm{V}=0.02$ between these values reflects the fact that the Sun is above the Galactic midplane and the main equatorial concentration of Galactic dust. 
Given the estimate of this offset as 15 kpc (Gontcharov 2008, 2011, 2012c), obviously, this extinction $\Delta A_\mathrm{V}=0.02$ arises in the dust layer below the Sun, 
within approximately $-30<|Z|<0$ pc. This alone does not allow the extinction within, say, 30 pc of the Sun to be deemed negligible, although, admittedly, the difference in 
total extinction toward the SGP and NGP is so small that so far it has been noticed only in some maps and models: $\Delta A_\mathrm{V}=0.013$, 0.010, 0.013, 0.011, 0.016, and 0.012
according to SFD98, DCL03, AM05, SF11, MF15, and TGE, respectively (see Fig. 8).

Note that the adopted correction influenced the extinction estimates predominantly by reducing them (i) near the Sun in our 3D maps, since in the range $2.9<D<5.9$, i.e. 
$38<R<150$ pc, the correction is negative, and (ii) at high latitudes in our 2D map, since we used dwarfs in the range $450<|Z|<2500$ pc, i.e., predominantly $900<R<2500$ pc, 
or $9.8<D<12$, where a negative correction prevails.

For the convenience of their use and interpolation, for our maps we adopted a uniform coordinate grid with a step of 20 arcmin for the 3D maps (given the influence of the cosine 
of the latitude on the longitude step) and 6.1 arcmin for the 2D map (coincides with the SFD98 resolution for the convenience of our comparison).

The coordinate grid of our maps was chosen so that for each latitude there is a reading at longitude $l=180^{\circ}$, while the remaining grid points for a given latitude are 
located symmetrically relative to the reading at $l=180^{\circ}$. Such a grid is devoid of readings near $l=0^{\circ}$, which, as the previous maps show, are least
useful due to the large gradients of all the quantities under consideration toward the Galactic center.

Although the step of the coordinate grid of our maps is fixed, to optimize the number of stars being used in the cells of the 3D maps, we took the size of the averaging window 
across the line of sight and, accordingly, the tangential resolution of the maps to be dependent on $R$. For a given $R$ we adopt the lower of the resolutions of 20 arcmin or 
3.55 pc (the latter corresponds to 6.1 arcmin at a distance of 2 kpc). Thus, at a distance from 50 to 2000 pc from the Sun the resolution changes from 4.06 degrees to 20 arcmin, or
from 3.55 to 11.64 pc. The adopted radial resolution of 50 pc is much poorer than the tangential resolution due to the great uncertainty in $R$. Future, more accurate parallaxes 
for more complete Gaia samples will probably allow the radial resolution of the maps to be increased.

Some maps of other authors have a higher angular (3.4 arcmin in GSZ19) or spatial (25 pc in LVB22) resolution. However, near the Sun or far from the Galactic midplane this leads 
to unreliable results due to the shortage of stars in spatial cells. For example, because of the insufficient number of stars in cells, GSZ19 gives an erroneous zero reddening 
for many cells far from the Galactic midplane (see the discussion in Gontcharov et al. (2022) and Fig. 8 in the Section``3D Maps'').

\begin{figure*}
\includegraphics{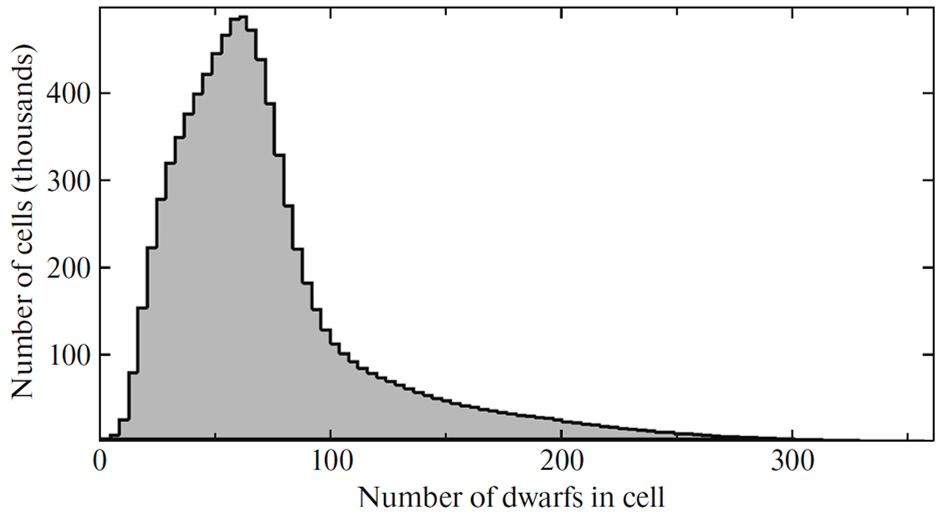}
\caption{The distribution of spatial cells of our 3D map in number of dwarfs used.
}
\label{histo}
\end{figure*}

Our 2D map contains 3\,095\,841 cells in the sky, while the 3D AV map contains 8\,724\,902 spatial cells. The distribution of spatial cells of the 3D map in number of dwarfs used 
is shown in Fig. 3.

In each cell the individual extinctions are averaged. The 2D map is the result of such averaging for stars with $|Z|>450$ pc and $R>2000$ pc and, accordingly, for latitudes 
$|b|>\arcsin(450/2000)\approx13^{\circ}$. For the 3D $A_\mathrm{V}$ and $A_\mathrm{G}$ maps, after the averaging in the previously described windows, there are numerous cases
where on one line of sight the extinction decreases with distance. This is the result of dust medium fluctuations and errors in the distance and extinction estimates. 
To eliminate this effect, we adjust the extinction estimates on each line of sight by increasing or decreasing the adjacent estimates by iterations in such a way that, 
as a result, the extinction did not decrease with distance. At the same time, the extinction estimates from our 3D maps at $|Z|=450$ pc agree with the estimates from our 2D map. 
In fact, the final curve of extinction growth with distance for a line of sight is the most probable non-decreasing curve passing among the individual extinction estimates on
this line of sight. This adjustment requires tens and hundreds of iterations and, therefore, is very demanding to computational resources. Such adjustment is performed when 
producing any up-to-date reddening/extinction map (see, e.g., GSZ19 and GCY21).

The 3D $A_\mathrm{G}/A_\mathrm{V}$ and differential extinction maps are calculated from the AV and AG maps.

The accuracy of our maps can be estimated from the following considerations. As follows from Fig. 3, to calculate the extinction in a spatial cell, as a rule, we used several 
tens of dwarfs. The statistics for the cells of the 2D map is approximately the same. At a random error of the individual extinction $\sigma(A_\mathrm{V})=0.20$ for a typical 
dwarf with a magnitude $G=18$ being considered by us, the random error of the result in 94\% of the cells is $\sigma(A_\mathrm{V})<0.04$. Note that the systematic accuracy of 
the AKQ22 extinctions used is fairly high, given that we found no its significant manifestations in the dependences on any parameters, except for the previously mentioned sinusoidal
dependence of the extinction on $R$. However, the uncertainty in the individual $R$, the surface gravity (via the dwarf selection criterion \verb"logg50" $>3.95$), and the systematic 
uncertainty equal to the applied correction for the empirical sinusoidal dependence of the extinction on R (since the cause of this systematics is unclear) reaching 0.04 mag also 
contribute to the total uncertainty of the result. Because of the increase in the cell size with $R$, the number of dwarfs in a cell decreases with $R$ rather slowly, while the random
extinction error is proportional to the square root of the number of dwarfs. In addition, the uncertainty in the individual $R$ within 2.5 kpc of the Sun increases with $R$ slowly. 
Moreover, the sinusoidal dependence of the extinction on $R$ does not depend on $R$ itself. Therefore, as a result, it can be guaranteed that the total uncertainty in the extinction 
estimates for extended objects in our maps everywhere does not exceed $\sigma(A_\mathrm{V})=0.06$. This precision is higher than the previously mentioned typical accuracy of the
best up-to-date maps $\sigma(A_\mathrm{V})=0.08$ due to the higher accuracy of the AKQ22 extinctions used and, to a lesser degree, due to the optimization of the cell size.

To estimate the accuracy of the predictions of our maps for individual stars and other point sources, we need to take into account the natural dust medium fluctuations, 
which range from $\sigma(A_\mathrm{V})=0.06$ near the Galactic poles to $\sigma(A_\mathrm{V})=0.33$ near the Galactic equator (and even higher in some small regions),
as determined by Gontcharov et al. (2022) from the standard deviation of the extinction estimates from AKQ22. Consequently, the accuracy of the predictions of our maps for a 
point object ranges from $\sigma(A_\mathrm{V})=(0.06^2+0.06^2)^{0.5}=0.08$ near the Galactic poles to $\sigma(A_\mathrm{V})=(0.06^2+0.33^2)^{0.5}=0.34$ near the Galactic equator. 
We fitted this dependence on latitude by a polynomial, $\sigma(A_\mathrm{V})=-4.0\cdot10^{-7}\cdot|b|^3+1.0\cdot10^{-4}\cdot|b|^2-0.0086|b|+0.34$.
Thus, at high latitudes, approximately at $|b|>60^{\circ}$, the predictions of our maps even for individual stars are more accurate than the predictions for the same stars from
AKQ22 with a typical uncertainty $\sigma(A_\mathrm{V})\approx0.18$. At lower latitudes the estimates of our and other similar reddening/extinction maps and models have a
practical application only for stars without individual estimates\footnote{Within 2 kpc of the Sun there are several billion such stars, or 99\% of all stars, including nearly 
400 million (about 80\%) stars from Gaia DR3).} or for extended objects (more than 10000 galaxies, star clusters, various clouds, and other objects).

\begin{table*}
 \centering
\def\baselinestretch{1}\normalsize\normalsize
\caption[]{Our 2D $A_\mathrm{V}$ maps (completely given in electronic form).
}
\label{solution2d}
\begin{tabular}[c]{ccccccc}
\hline
\noalign{\smallskip}
$l$       &    $b$    & Minimum $R$ & Window      & Number of dwarfs & $A_\mathrm{V}$ \\
(degs) & (degs) & (parsecs)       & (degs) &                &    \\
\hline
\noalign{\smallskip}
180.0000 & $-89.9750$ & 450.0  & 0.102  &  19 & 0.138 \\
180.0000 & $-89.8733$ & 450.0  & 0.102  &  17 & 0.116 \\
134.0141 & $-89.8733$ & 450.0  & 0.102  &   9 & 0.087 \\
 88.0281 & $-89.8733$ & 450.0  & 0.102  &  10 & 0.101 \\
 42.0422 & $-89.8733$ & 450.0  & 0.102  &  15 & 0.126 \\
225.9859 & $-89.8733$ & 450.0  & 0.102  &  13 & 0.125 \\
271.9719 & $-89.8733$ & 450.0  & 0.102  &  13 & 0.154 \\
317.9578 & $-89.8733$ & 450.0  & 0.102  &  22 & 0.168 \\
\ldots    &   \ldots    & \ldots & \ldots  & \ldots & \ldots & \ldots \\
\hline
\end{tabular}
\end{table*}


\begin{figure*}
\includegraphics{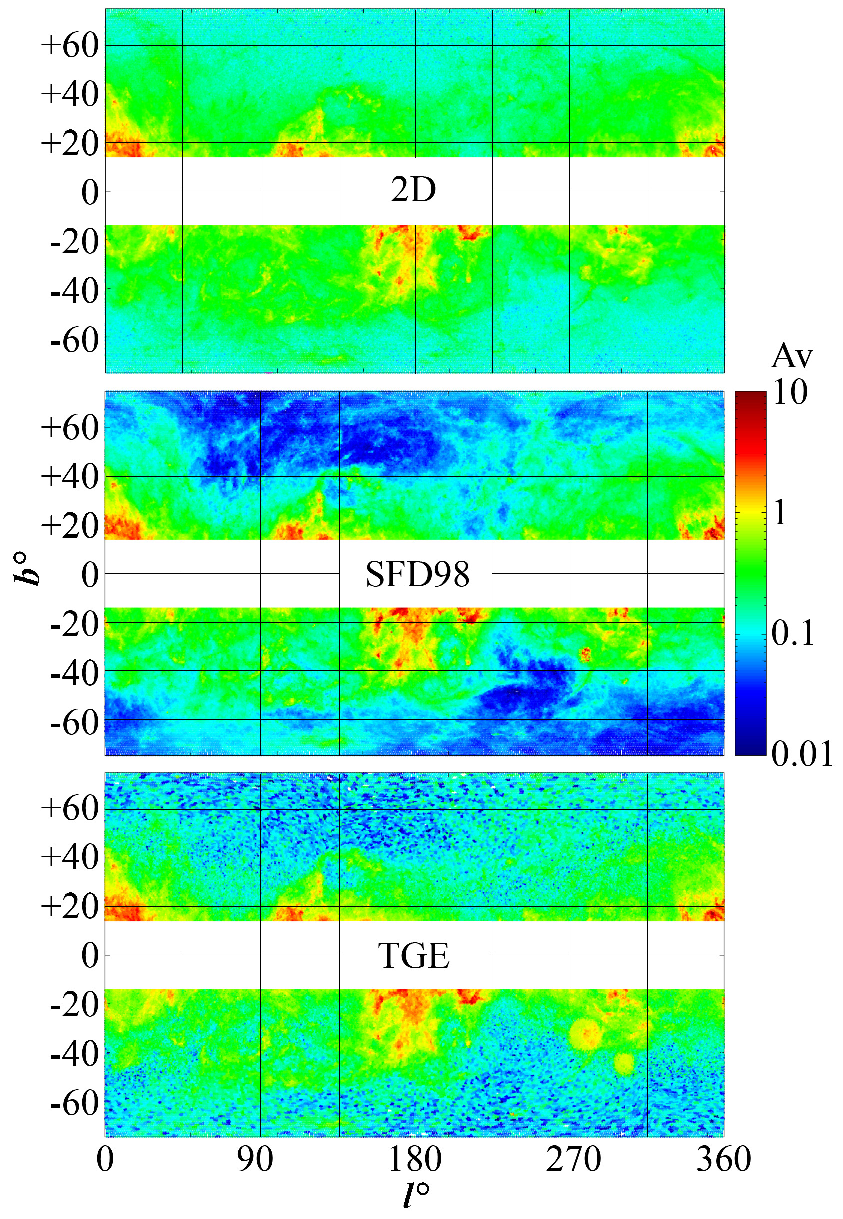}
\caption{Variations in extinction $A_\mathrm{V}$ over the sky in accordance with our 2D map and the SFD98 and TGE maps.
}
\label{2dlb}
\end{figure*}

\begin{figure*}
\includegraphics{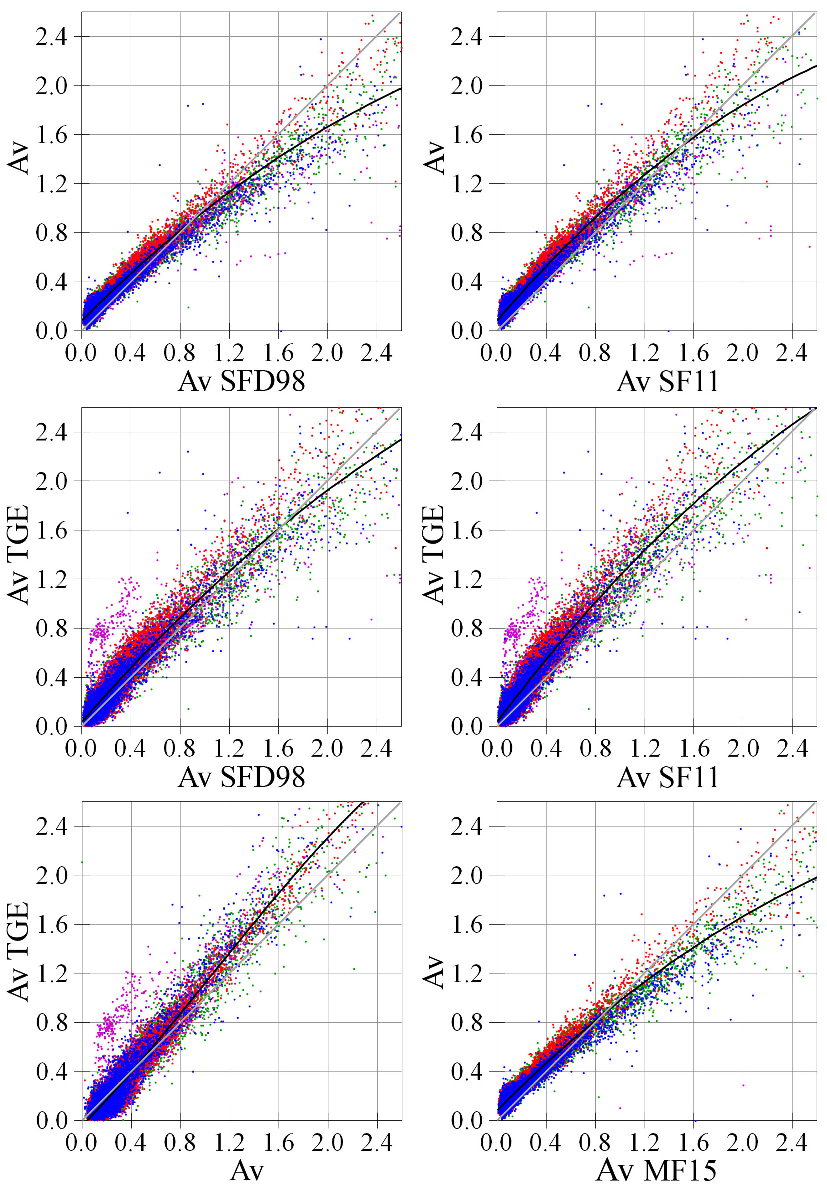}
\caption{Extinction $A_\mathrm{V}$ from our 2D map in comparison with its estimates from the 2D SFD98, SF11, TGE, and MF15 maps for randomly selected 32\,000 lines of sight in the 
first (red symbols), second (green symbols), third (blue symbols), and fourth (violet symbols) Galactic quadrants. The estimates for the first and third quadrants are brought to 
the foreground. The gray straight lines indicate the one-to-one relation; the black curves indicate the best fit to the relation between the estimates by a cubic polynomial.
}
\label{2d}
\end{figure*}

\begin{figure*}
\includegraphics{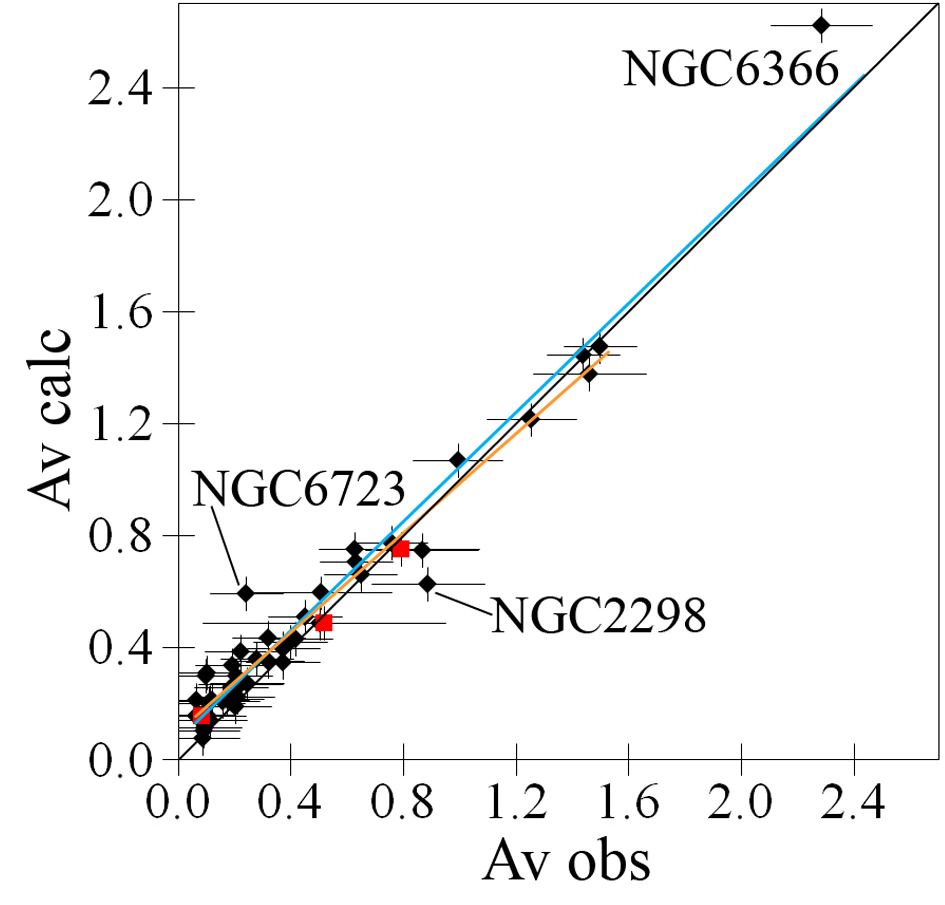}
\caption{Estimates of $A_\mathrm{V\,obs}$ for Galactic globular clusters from the literature in comparison with the predictions of $A_\mathrm{V\,calc}$ from our 2D map (black symbols). 
The estimates from Clementini et al. (2022) for NGC\,288, NGC\,5139, and IC\,4499 are marked by the red squares. The black straight line indicates the one-to-one relation; 
the blue and orange straight lines indicate the linear approximation of the relation between the observations and predictions by the least-squares method with and without including
NGC\,6366, respectively.
}
\label{gc}
\end{figure*}

\section*{2D MAP}
\label{2dmap}

Table 1 presents our 2D map. Figure 4 shows the variations in extinction $A_\mathrm{V}$ over the sky in accordance with our 2D map and the 2D SFD98 and TGE maps 
(for TGE we adopted the optimal angular resolution recommended by the authors). For the convenience of our comparison, we excluded the low latitudes where there are no our 
estimates. It can be seen that at middle latitudes all maps agree with one another, demonstrating primarily the well-known cloud complexes of the Gould Belt in the two opposite
regions with coordinates $l\approx0^{\circ}$, $b\approx+20^{\circ}$ and $l\approx180^{\circ}$, $b\approx-20^{\circ}$ that were discussed, for example, by Dame et al. (2001), 
Gontcharov (2009, 2012b), and Gontcharov et al. (2022). At high latitudes Fig. 4 shows that our estimates, as a rule, are several times higher than the SFD98 estimates, whereas 
the TGE estimates are intermediate. This can also be seen in the estimates of the total Galactic extinction through the entire dust layer toward the Galactic poles: 
$A_\mathrm{V}=0.12$, 0.05, 0.07 from our 2D, SFD98, and TGE maps, respectively.

Figure 5 compares the estimates from our 2D map with the estimates from the SFD98, SF11, MF15, and TGE maps for 32\,000 randomly selected $6.1\times6.1$ arcmin fields 
(different colors indicate the estimates in different Galactic quadrants). Note that the TGE estimates clearly deviating from the bisector (the violet symbols, i.e., the estimates 
in the fourth Galactic quadrant) refer to the sky regions with the Magellanic Clouds, where, as expected, it is difficult to take into account the extinction inside these clouds.

Note the linear relationship between the SFD98 and SF11 estimates due to the SF11 production method and the previously known closeness of the SFD98 and MF15 estimates.

TGE used the same data as we did, but processed them by a different method. This explains the good agreement of our estimates with the TGE ones in systematic terms. 
Figure 5 shows an approximately linear relationship between the TGE and our estimates, although both maps show significantly non-linear relationships with the SFD98, SF11, and MF15 
maps obtained from the IR emission. Thus, both TGE and our map confirm the underestimation of low reddenings/extinctions by the SFD98, SF11, and MF15 maps known from previous 
studies (for a review, see Gontcharov 2016b) and, possibly, the overestimation of high ones. However, for high extinctions this effect is evened out by the fact that both our map and 
TGE may underestimate high extinctions, because they possibly do not pierce through the entire Galactic dust layer, losing the completeness of the sample in distant dusty regions.

The underestimation of low reddenings/extinctions by the SFD98, MF15, and SF11 maps due to their systematic errors, primarily the IR emission--reddening calibration errors, 
requires a separate study. Our results show that the AKQ22 data and the subsequent Gaia results are a suitable material for such a study. In this paper we will only note the
dependence of the difference between our estimates and the estimates of these three maps on Galactic longitude. To emphasize this dependence, in Fig. 5 we brought the estimates 
for the first and third quadrants to the foreground (red and blue symbols) and the equally numerous estimates for the second and fourth quadrants to the background. It can be
seen that the deviation of the estimates from the bisector depends on $l$ and is extreme in the first and third quadrants. This is possibly because in these quadrants the extinction, 
as a rule, is maximal and minimal, respectively.

Although our and TGE estimates agree in systematic terms, the random scatter of their differences is great, especially for low extinctions, as can be seen from Fig. 5: 
for example, at $|b|>45^{\circ}$ the median standard deviation of the $A_\mathrm{V}$ differences is 0.06. For low extinctions (and high latitudes) the SFD98, SF11, and MF15 
estimates agree with ours in random terms much better (at $|b|>45^{\circ}$ the median standard deviation is 0.03) than with the TGE estimates (at $|b|>45^{\circ}$ the median 
standard deviation is 0.06). Some methodological shortcomings of TGE are a possible reason. For high extinctions (and middle latitudes) the situation is completely different:
our estimates agree with TGE much better (at $|b|<20^{\circ}$ the median standard deviation is 0.2) than do both with SFD98, SF11, and MF15 (at $|b|<20^{\circ}$ the median 
standard deviation is about 0.4). A possible reason is a poorer angular resolution of our map and TGE than that of the remaining ones and, consequently, a stronger smoothing 
of the natural medium fluctuations.

Note also that our estimates agree excellently with those from SF11 in both random and systematic terms if the constant extinction $\Delta A_\mathrm{V}=0.08$ is added to the latter. 
This probably reflects the systematic underestimation of low extinctions in SF11 and approximately the same one in SFD98 and MF15.

To estimate the accuracy of the predictions of our 2D map, we compared them in Table 2 and Fig. 6 with the most accurate extinction estimates for Galactic globular clusters 
with $|b|>13^{\circ}$ obtained (i) by Clementini et al. (2022) from the analysis of RR~Lyrae variables in the clusters NGC\,288, NGC\,5139, and IC\,4499 (red squares) and (ii) by
different authors from the comparison of photometry and theoretical isochrones on the color--magnitude diagrams and by other methods independent of the estimates from the 
extinction maps and models (black diamonds). In the latter case, we considered 47 known globular clusters within 30 kpc of the Sun (for more distant clusters the estimates from the
literature are inaccurate). The distance estimates for the clusters were taken from Baumgardt and Vasiliev (2021), except for the clusters Terzan 7 and Terzan 8 with the estimates 
from Harris (1996). We took the reddening/extinction estimates for the clusters from the studies by Gontcharov et al. (2019, 2020, 2021, 2023a, 2023b), Dotter et al. (2011), 
Bellazzini et al. (2002), Koch and McWilliam (2014), Hamrick et al. (2021), Recio-Blanco et al. (2005), Wagner-Kaiser et al. (2016), and Wagner-Kaiser et al. (2017) 
with the preference precisely in this order if there are estimates for a cluster in different publications. For these estimates we adopted the most realistic (in our view) 
estimate of the uncertainty by taking into account the extinction variations in the cluster field based on data from Bonatto et al. (2013), but no less than $\sigma(A_\mathrm{V})=0.13$. 
For the estimates from Clementini et al. (2022) we adopted the uncertainties declared by the authors. All of the reddening or extinction estimates used by us were converted to the 
estimates of the extinction $A_\mathrm{V}$ using the extinction law from Cardelli et al. (1989) with the nominal extinction-to-reddening ratio 
$R_\mathrm{V}\equiv A_\mathrm{V}/E(B-V)=3.1$, given that the observed ratio $A_\mathrm{V}/E(B-V)$ depends on the spectral energy distribution of an unreddened star. Accordingly, 
for globular clusters with a predominance of comparatively cool and metal-poor stars we adopted the median ratio $A_\mathrm{V}/E(B-V)=3.3$ from Casagrande and VandenBerg (2014).

\begin{table*}
 \centering
\def\baselinestretch{1}\normalsize\footnotesize
\caption[]{Estimates of $A_\mathrm{V\,obs}$ for Galactic globular clusters with $|b|>13^{\circ}$ from the literature in comparison with the predictions of $A_\mathrm{V\,calc}$ 
from our 2D map.
}
\label{gctab}
\begin{tabular}[c]{lrrrccl}
\hline
\noalign{\smallskip}
Имя         & $R$~(kpc)  & $l$~(deg) & $b$~(deg)  & $A_\mathrm{V\,calc}$  & $A_\mathrm{V\,obs}$ & Reference \\
\hline
\noalign{\smallskip}
IC4499      &       18.9 &        307.35 & $-20.47$       &                  0.75 & $0.63\pm0.13$       & Dotter et al. (2011) \\
IC4499      &       18.9 &        307.35 & $-20.47$       &                  0.75 & $0.79\pm0.28$       & Clementini et al. (2022) \\
NGC104      &        4.5 &        305.90 & $-44.89$       &                  0.16 & $0.11\pm0.13$       & Wagner-Kaiser et al. (2017) \\
NGC288      &        9.0 &        151.28 & $-89.38$       &                  0.16 & $0.08\pm0.16$       & Gontcharov et al. (2021) \\
NGC288      &        9.0 &        151.28 & $-89.38$       &                  0.16 & $0.08\pm0.09$       & Clementini et al. (2022) \\
NGC362      &        8.8 &        301.54 & $-46.25$       &                  0.21 & $0.11\pm0.13$       & Gontcharov et al. (2021) \\
NGC1261     &       16.4 &        270.54 & $-52.12$       &                  0.10 & $0.09\pm0.13$       & Wagner-Kaiser et al. (2016) \\
NGC1851     &       12.0 &        244.51 & $-35.04$       &                  0.21 & $0.16\pm0.13$       & Wagner-Kaiser et al. (2017) \\
NGC1904     &       13.1 &        227.23 & $-29.35$       &                  0.31 & $0.10\pm0.13$       & Hamrick et al. (2021) \\
NGC2298     &        9.8 &        245.63 & $-16.01$       &                  0.63 & $0.89\pm0.20$       & Wagner-Kaiser et al. (2017) \\
NGC4147     &       18.5 &        252.85 & $+77.19$       &                  0.14 & $0.11\pm0.13$       & Wagner-Kaiser et al. (2017) \\
NGC4590     &       10.4 &        299.63 & $+36.05$       &                  0.27 & $0.22\pm0.15$       & Wagner-Kaiser et al. (2017) \\
NGC5024     &       18.5 &        332.96 & $+79.76$       &                  0.12 & $0.09\pm0.13$       & Wagner-Kaiser et al. (2016) \\
NGC5053     &       17.5 &        335.70 & $+78.95$       &                  0.16 & $0.07\pm0.13$       & Wagner-Kaiser et al. (2017) \\
NGC5139     &        5.4 &        309.10 & $+14.97$       &                  0.49 & $0.50\pm0.13$       & Wagner-Kaiser et al. (2017) \\
NGC5139     &        5.4 &        309.10 & $+14.97$       &                  0.49 & $0.52\pm0.43$       & Clementini et al. (2022) \\
NGC5272     &       10.2 &         42.22 & $+78.71$       &                  0.08 & $0.09\pm0.13$       & Wagner-Kaiser et al. (2016) \\
NGC5466     &       16.1 &         42.15 & $+73.59$       &                  0.16 & $0.06\pm0.13$       & Wagner-Kaiser et al. (2017) \\
NGC5634     &       26.0 &        342.21 & $+49.26$       &                  0.30 & $0.20\pm0.13$       & Bellazzini et al. (2002) \\
NGC5897     &       12.5 &        342.94 & $+30.29$       &                  0.42 & $0.40\pm0.13$       & Koch and McWilliam (2014) \\
NGC5904     &        7.5 &          3.86 & $+46.80$       &                  0.19 & $0.20\pm0.13$       & Gontcharov et al. (2019) \\
NGC5986     &       10.5 &        337.02 & $+13.27$       &                  1.07 & $0.99\pm0.16$       & Wagner-Kaiser et al. (2017) \\
NGC6093     &       10.3 &        352.67 & $+19.46$       &                  0.78 & $0.76\pm0.13$       & Wagner-Kaiser et al. (2017) \\
NGC6101     &       14.4 &        317.75 & $-15.82$       &                  0.43 & $0.42\pm0.13$       & Wagner-Kaiser et al. (2016) \\
NGC6121     &        1.9 &        350.97 & $+15.97$       &                  1.45 & $1.44\pm0.13$       & Wagner-Kaiser et al. (2017) \\
NGC6144     &        8.2 &        351.93 & $+15.70$       &                  1.48 & $1.50\pm0.13$       & Wagner-Kaiser et al. (2017) \\
NGC6171     &        5.6 &          3.37 & $+23.01$       &                  1.38 & $1.46\pm0.20$       & Wagner-Kaiser et al. (2016) \\
NGC6205     &        7.4 &         59.01 & $+40.91$       &                  0.21 & $0.12\pm0.13$       & Gontcharov et al. (2020) \\
NGC6218     &        5.1 &         15.72 & $+26.31$       &                  0.71 & $0.63\pm0.13$       & Gontcharov et al. (2021) \\
NGC6229     &       30.1 &         73.64 & $+40.31$       &                  0.21 & $0.06\pm0.13$       & Recio-Blanco et al. (2005) \\
NGC6254     &        5.1 &         15.14 & $+23.08$       &                  0.75 & $0.86\pm0.20$       & Wagner-Kaiser et al. (2016) \\
NGC6341     &        8.5 &         68.34 & $+34.86$       &                  0.30 & $0.10\pm0.13$       & Wagner-Kaiser et al. (2016) \\
NGC6362     &        7.6 &        325.55 & $-17.57$       &                  0.34 & $0.19\pm0.13$       & Gontcharov et al. (2023a) \\
NGC6366     &        3.4 &         18.41 & $+16.04$       &                  2.62 & $2.28\pm0.18$       & Wagner-Kaiser et al. (2016) \\
NGC6426     &       20.7 &         28.09 & $+16.23$       &                  1.21 & $1.25\pm0.16$       & Dotter et al. (2011) \\
NGC6584     &       13.6 &        342.14 & $-16.41$       &                  0.44 & $0.32\pm0.13$       & Wagner-Kaiser et al. (2016) \\
NGC6715     &       26.3 &          5.61 & $-14.09$       &                  0.60 & $0.51\pm0.25$       & Wagner-Kaiser et al. (2017) \\
NGC6723     &        8.3 &          0.07 & $-17.30$       &                  0.59 & $0.24\pm0.13$       & Gontcharov et al. (2023a) \\
NGC6752     &        4.1 &        336.49 & $-25.63$       &                  0.27 & $0.24\pm0.13$       & Wagner-Kaiser et al. (2017) \\
NGC6809     &        5.3 &          8.79 & $-23.27$       &                  0.35 & $0.37\pm0.13$       & Gontcharov et al. (2023b) \\
NGC6864     &       20.5 &         20.30 & $-25.75$       &                  0.66 & $0.65\pm0.13$       & Recio-Blanco et al. (2005) \\
NGC6934     &       15.7 &         52.10 & $-18.89$       &                  0.40 & $0.37\pm0.13$       & Wagner-Kaiser et al. (2016) \\
NGC6981     &       16.7 &         35.16 & $-32.68$       &                  0.26 & $0.19\pm0.13$       & Wagner-Kaiser et al. (2016) \\
NGC7078     &       10.7 &         65.01 & $-27.31$       &                  0.35 & $0.32\pm0.13$       & Wagner-Kaiser et al. (2016) \\
NGC7089     &       11.7 &         53.37 & $-35.77$       &                  0.22 & $0.21\pm0.13$       & Wagner-Kaiser et al. (2017) \\
NGC7099     &        8.5 &         27.18 & $-46.84$       &                  0.22 & $0.17\pm0.13$       & Wagner-Kaiser et al. (2016) \\
Palomar5    &       21.9 &          0.84 & $+45.86$       &                  0.36 & $0.28\pm0.13$       & Dotter et al. (2011) \\
Palomar12   &       18.5 &         30.51 & $-47.68$       &                  0.20 & $0.11\pm0.13$       & Wagner-Kaiser et al. (2017) \\
Terzan7     &       22.8 &          3.39 & $-20.07$       &                  0.39 & $0.22\pm0.13$       & Wagner-Kaiser et al. (2017) \\
Terzan8     &       26.3 &          5.76 & $-24.56$       &                  0.51 & $0.45\pm0.13$       & Wagner-Kaiser et al. (2017) \\
\hline
\end{tabular}
\end{table*}

Figure 6 shows good agreement between the predictions of our map (with the uncertainty $\sigma(A_\mathrm{V})=0.06$) and the estimates from the literature. The blue and orange 
straight lines in Fig. 6 indicate the linear fit to the relation between the observations and predictions by the least-squares method with and without including NGC\,6366, 
respectively. It can be seen that for low extinctions our estimates may be systematically higher than the estimates from the literature, though within a few hundredths of a magnitude.

The deviation of the clusters in Fig. 6 from the bisector can be explained, in particular, by the spatial variations of the extinction law or the ignored extinction variations 
in the cluster field. For example, NGC\,6366 is only 16 arcmin away from the bright fifth-magnitude star 47~Oph, while the cluster radius is at least 9.5 arcmin (Bica et al. 2019).
Under the influence of 47~Oph, in NGC\,6366 there arise a strong gradient of the zero point of photometric measurements and other sources of systematic errors in the 
extinction/reddening estimates described by Anderson et al. (2008) and detected by Bonatto et al. (2013) as a large extinction gradient in the NGC\,6366 field. 
Similarly, the disagreement of the estimates for NGC\,6723 can be explained by the fact that this cluster is projected onto the edge of the Corona Australis complex of clouds, 
causing a large extinction gradient in the NGC\,6723 field (Gontcharov et al. 2023a), while for NGC\,2298 Bonatto et al. (2013) found the maximum extinction gradient among all 
of the globular clusters investigated by them, although the direct reason for this is unclear.

\begin{table*}
 \centering
\def\baselinestretch{1}\normalsize\normalsize
\caption[]{Our 3D $A_\mathrm{V}$ map (completely given in electronic form).
}
\label{solution3d}
\begin{tabular}[c]{cccccc}
\hline
\noalign{\smallskip}
$l$       &    $b$    & $R$        & Window      & Number of dwarfs & $A_\mathrm{V}$  \\
(deg) & (deg) & (pc)  & (deg) &                &   (mag)    \\
\hline
\noalign{\smallskip}
180.0000 & $-89.6667$ & 500 &  0.4 &  18 &  0.129  \\
180.0000 & $-89.6667$ & 450 &  0.5 &  16 &  0.129  \\
180.0000 & $-89.6667$ & 400 &  0.5 &  20 &  0.129  \\
180.0000 & $-89.6667$ & 350 &  0.6 &  24 &  0.128 \\
180.0000 & $-89.6667$ & 300 &  0.7 &  17 &  0.128 \\
180.0000 & $-89.6667$ & 250 &  0.8 &  16 &  0.128 \\
180.0000 & $-89.6667$ & 200 &  1.0 &  25 &  0.128 \\
180.0000 & $-89.6667$ & 150 &  1.4 &  18 &  0.128 \\
180.0000 & $-89.6667$ & 100 &  2.0 &  16 &  0.078 \\
180.0000 & $-89.6667$ &  50 &  4.1 &  24 &  0.078 \\
\ldots   &   \ldots   & \ldots & \ldots & \ldots & \ldots \\
\hline
\end{tabular}
\end{table*}


\begin{figure*}
\includegraphics{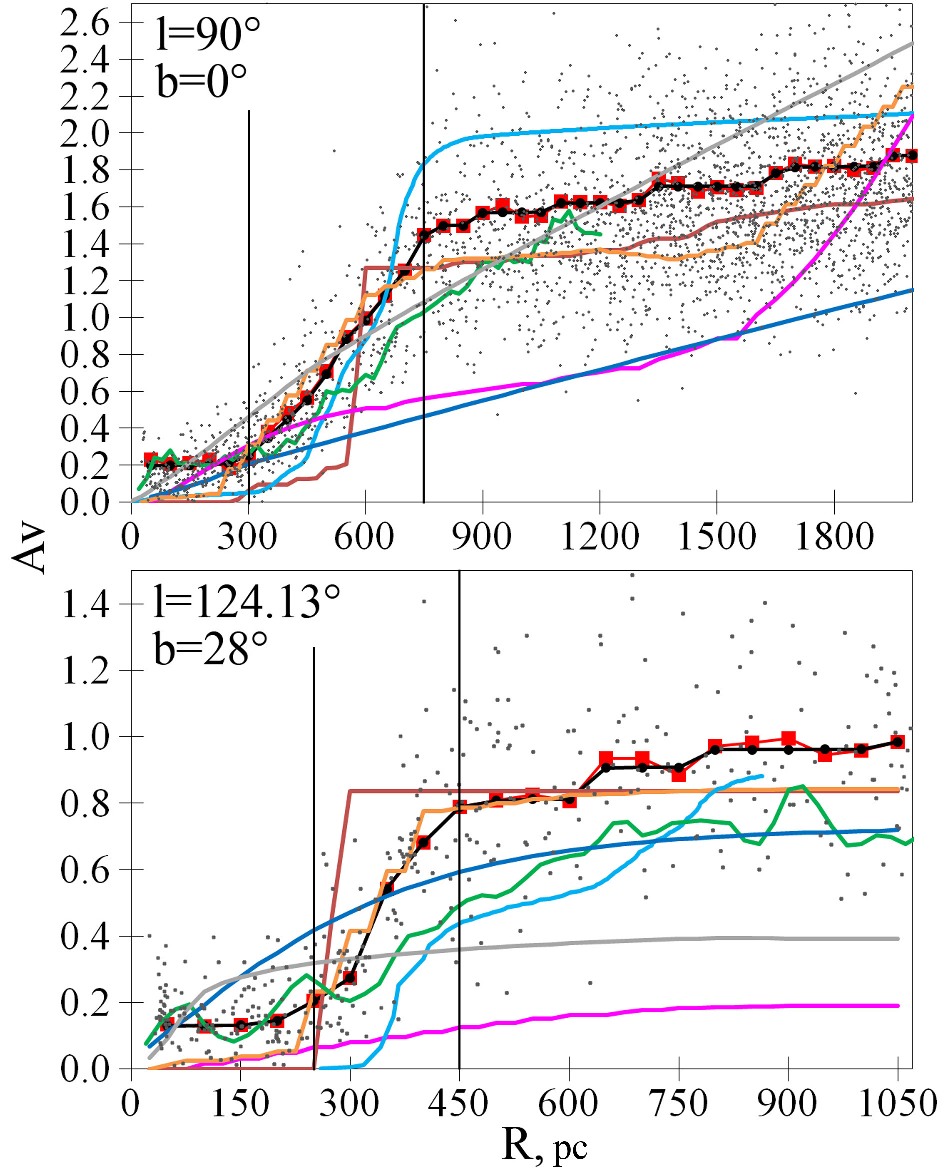}
\caption{Extinction $A_\mathrm{V}$ versus $R$ for two lines of sight toward the Cygnus ($l=90^{\circ}$, $b=0^{\circ}$) and Polaris Flare ($l=124.13^{\circ}$, $b=+28^{\circ}$) clouds. 
The curves of different colors indicate the estimates of different maps and models: our map before (red) and after (black) adjustment, DCL03 (blue), AL05 (magenta), G17 (green), 
GSZ19 (brown), GCY21 (orange), LVB22 (cyan), and GMS22 (gray). The individual dots indicate the AKQ22 dwarfs used by us. The vertical straight lines indicate the range of
distances where $A_\mathrm{V}$ grows significantly.
}
\label{sightlines}
\end{figure*}

\begin{figure*}
\includegraphics{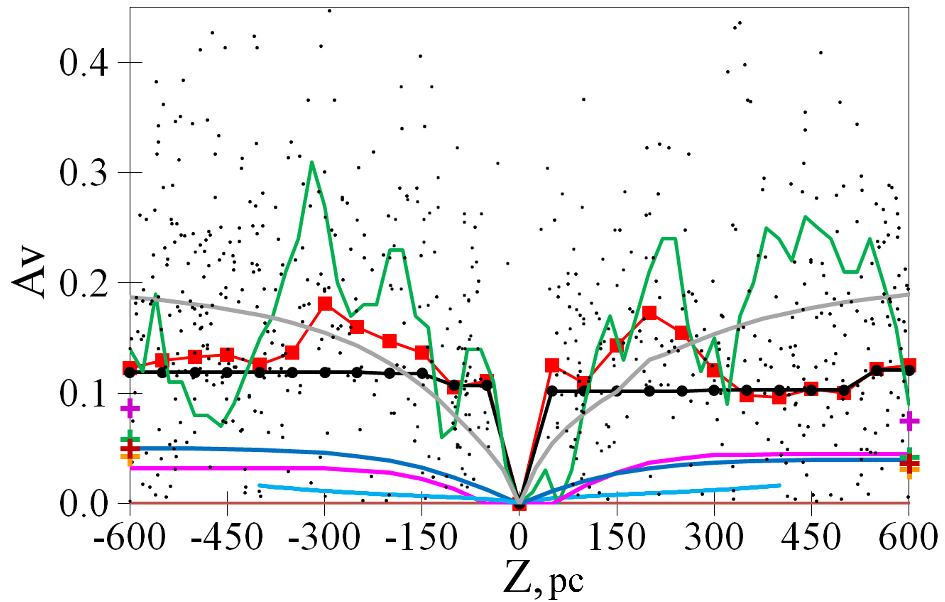}
\caption{Extinction $A_\mathrm{V}$ versus $Z$ toward the Galactic poles. The curves of different colors indicate the estimates of different maps and models: our map before (red) 
and after (black) adjustment, DCL03 (blue), AL05 (magenta), G17 (green), GSZ19 (brown), LVB22 (cyan), and GMS22 (gray). The color crosses at $|Z|=600$ pc indicate the estimates 
from the 2D maps: SFD98 (brown), SF11 (orange),MF15 (green), and TGE (violet). The individual dots indicate the AKQ22 dwarfs used by us.
}
\label{sgpngp_}
\end{figure*}

\begin{figure*}
\includegraphics{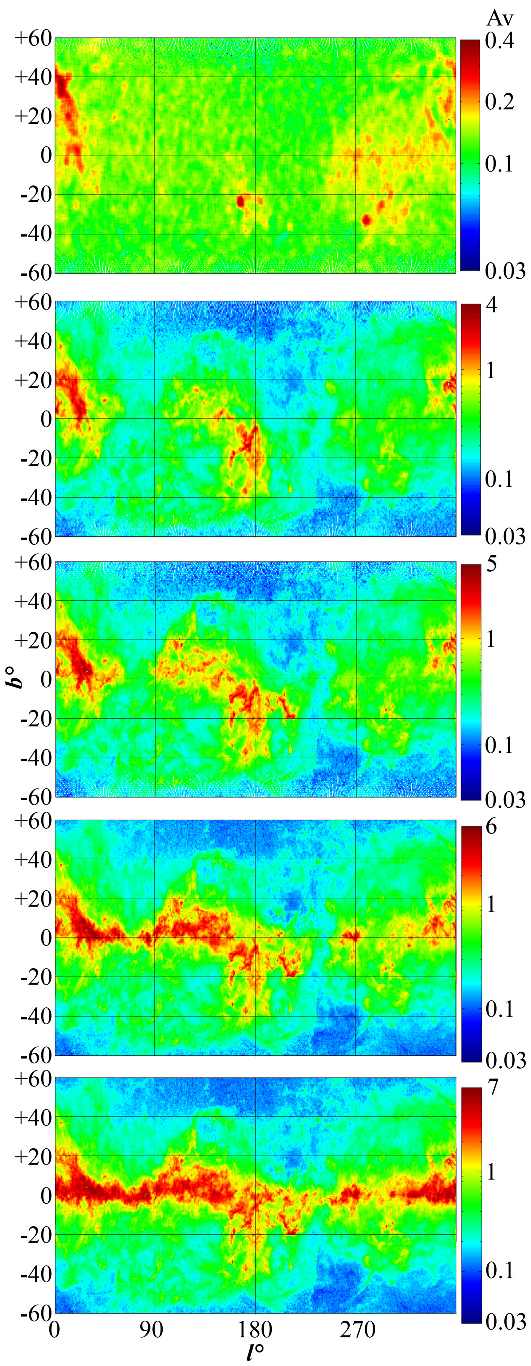}
\caption{Variations in extinction $A_\mathrm{V}$ over the sky to (from top to bottom) $R=100$, 300, 500, 1000, and 1650 pc.
}
\label{slices}
\end{figure*}

\begin{figure*}
\includegraphics{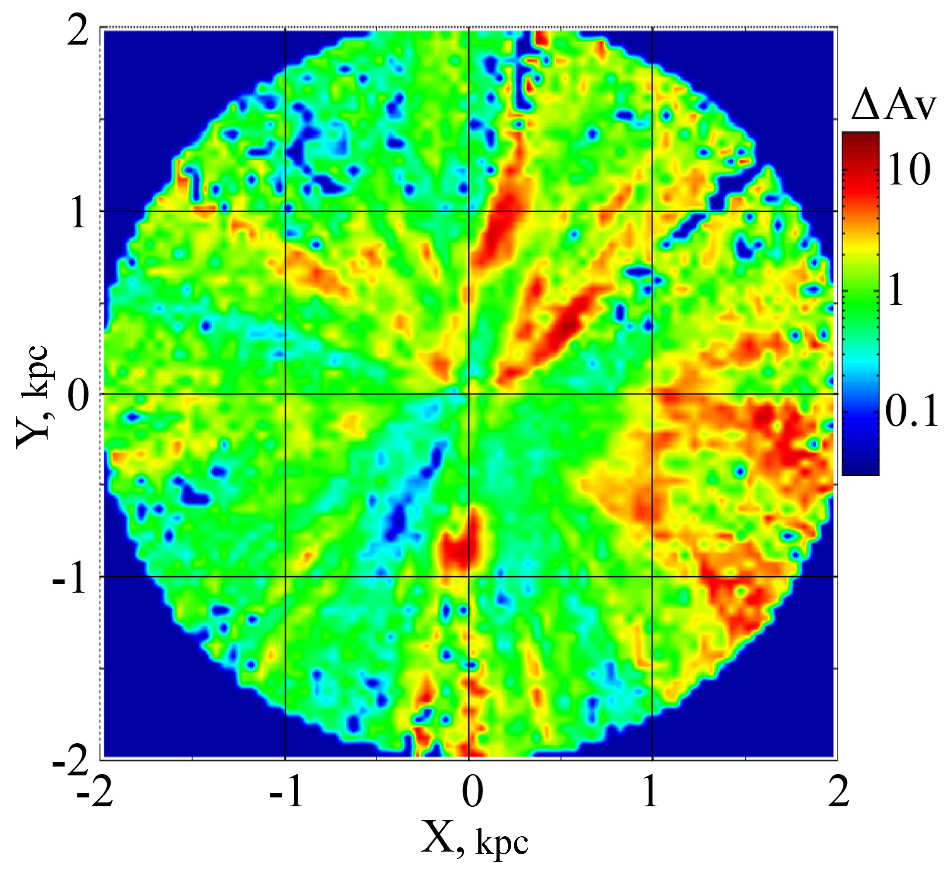}
\caption{Differential extinction along the line of sight (in magnitudes per kiloparsec) in the Galactic midplane based on the estimates from our map.
}
\label{z0}
\end{figure*}

\begin{figure*}
\includegraphics{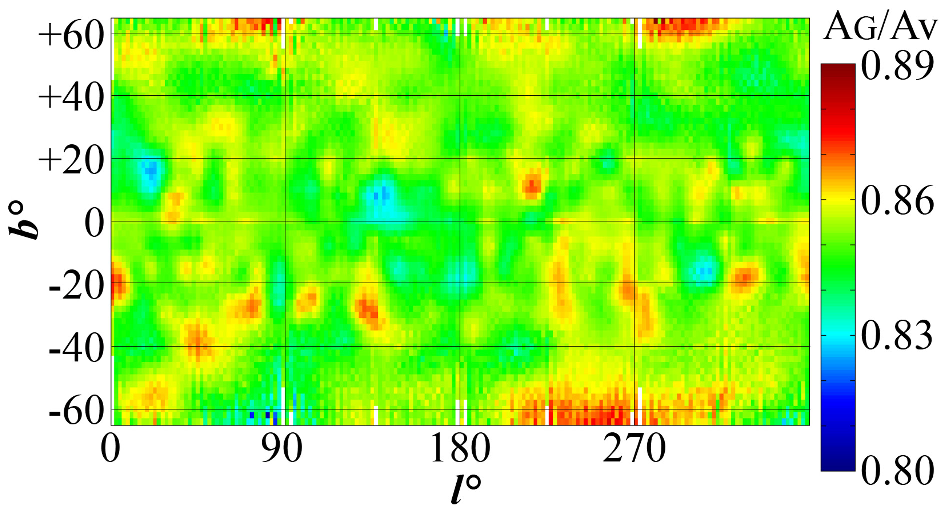}
\caption{$A_\mathrm{G}/A_\mathrm{V}$ variations over the sky within 800 pc of the Sun.
}
\label{agav}
\end{figure*}

\section*{3D MAPS}
\label{3dmap}

Table 3 presents our 3D $A_\mathrm{V}$ map. The 3D $A_\mathrm{G}$ map is unlikely to be of interest in its own right due to the similarity of the $V$ and $G$ filters and was 
obtained for the sake of the ratio $A_\mathrm{G}/A_\mathrm{V}$, which may reflect the extinction law and is discussed below.

Figure 7 shows an example of the $A_\mathrm{V}$ variations with distance for two lines of sight toward well-known clouds at low (the Cygnus cloud) and middle (the Polaris Flare cloud) 
latitudes. Similarly, Fig. 8 shows the $A_\mathrm{V}$ variations with $Z$ toward the Galactic poles. It can be seen that our estimates reproduce well the original estimates 
from AKQ22, while the adjustment (black symbols) corrected successfully the cases of a decrease in $A_\mathrm{V}$ with distance (red symbols), which is particularly important 
near the Sun.

For comparison, Figs. 7 and 8 show the $A_\mathrm{V}$ variations from the estimates of different 3D maps and analytical models of spatial reddening/extinction variations: 
Gontcharov et al. (2022, hereafter GMS22) with the parameters derived from the AKQ22 estimates for giants, Am\^ores and L\'epine (2005, hereafter AL05),\footnote{http://www.galextin.org/},
DCL03, G17, GSZ19, GCY21, and LVB22. In addition, the color crosses in Fig. 8 indicate the estimates at $|Z|=600$ pc from the 2D maps. It can be seen that the DCL03 map was calibrated
from the SFD98 map: the brown crosses in Fig. 8 are located exactly at the ends of the blue curves. Note that the zero or very low estimates toward the poles and near the Sun are 
the result of systematic errors in the case of LVB22 and poor original data in the case of GSZ19, as discussed by Gontcharov and Mosenkov (2019, 2021a, 2021b) and in GMS22.
For the 2D maps their systematic errors may also be responsible for the comparatively low $A_\mathrm{V}$ estimates toward the poles in Fig. 8.

It should be remembered that these maps and models are based on different data sets. It is the large differences in the original estimates that led to an enormous difference 
between the estimates of the maps and models toward the poles in Fig. 8. Consequently, it seems to us that the only way to obtain reliable reddening/extinction estimates at 
high latitudes is to obtain numerous individual estimates for different test objects (stars, globular clusters, variable stars, etc.) by different methods.

All analytical models (DCL03, AL05, GMS22) smooth out strongly the $A_\mathrm{V}$ variations in both tangential and radial directions. Therefore, in the case of large $A_\mathrm{V}$
gradients, the models can greatly under- or overestimate $A_\mathrm{V}$ on a specific line of sight. Furthermore, in the DCL03 and AL05 models more attention is given to Galactic 
regions far from the Sun to the detriment of the nearest hundreds of parsecs. This all can be seen in Figs. 7 and 8.

Figure 7 shows comparatively good agreement of the estimates of the 3D G17, GSZ19, and GCY21 maps between themselves and with our estimates, whereas the LVB22 estimates deviate 
from the remaining ones. This agreement allows one to estimate the range of distances where there are clouds on given lines of sight: 300--750 pc for the Cygnus cloud and 250--450 pc
for the Polaris Flare cloud. These estimates are consistent with the universally accepted ones. Thus, the 3D map can serve as a material to estimate the characteristics of 
large dust clouds.

Figure 9 shows the extinction $A_\mathrm{V}$ from our 3D map as a function of Galactic coordinates for different $R$. It can be seen that our map reproduces successfully the 
previously known dust structures (for comparison, see, e.g., Fig. 2 from Dame et al. (2001), Figs. 1--3 from GSZ19, and Fig. 12 from AKQ22).
At the same time, in accordance with GMS22, the clouds of the Gould Belt (the most noticeable ones near $l\approx0^{\circ}$, $b\approx+20^{\circ}$ and 
$l\approx180^{\circ}$, $b\approx-20^{\circ}$) and the clouds of the Cepheus--Chamaeleon dust layer (the most noticeable ones near $l\approx135^{\circ}$, $b\approx+15^{\circ}$
and $l\approx300^{\circ}$, $b\approx-20^{\circ}$) make a major contribution in the ranges $100<R<500$ pc and $300<R<1000$ pc, respectively, while at $R>1000$ pc the
influence of the equatorial dust layer grows.

Figure 9 shows that the region $R<100$ pc, i.e., the Local Bubble, differs significantly from the more distant regions by comparatively small extinction gradients and a fairly 
uniform growth of the extinction with distance. As a result, even within 100 pc of the Sun there are almost no sky regions with $A_\mathrm{V}<0.1$, as can be seen from Fig. 9. 
At the same time, in many high-latitude sky regions the extinction remains at a level $A_\mathrm{V}\approx0.1$ at $R>100$ pc. Thus, in the Bubble the extinction grows, on average, 
by $A_\mathrm{V}/R\approx0.1/0.1=1$ mag per kpc, i.e., no less than in many more distant regions. This can be seen on the map of differential extinction $A_\mathrm{V}/R$ along line 
of sight (i.e., the spatial dust density) that we produced based on our 3D $A_\mathrm{V}$ map. This map is presented in Table 4. As an example, the differential extinction along 
line of sight (in mag per kpc) in the Galactic midplane is shown in Fig. 10. This figure can be compared with the analogous Fig. 3 from LVB22, and it can be concluded that this 
differential extinction map allows large dust clouds to be seen. The region of reduced differential extinction nearest to the Sun is seen in Fig. 10 in the third quadrant at a 
distance of more than 100 pc from the Sun and not around the Sun. These results agree completely with the description of the Local Bubble by Gontcharov and Mosenkov (2019) as a 
region of ordinary density but enhanced ionization of the medium, where, accordingly, there is no current star formation. This view of the Bubble once again casts doubt on all the 
low estimates of $A_\mathrm{V}<0.1$ at high latitudes shown, for example, in Fig. 8.

\begin{table*}
 \centering
\def\baselinestretch{1}\normalsize\normalsize
\caption[]{Our 3D map of differential extinction $A_\mathrm{V}/R$ along line of sight (completely given in electronic form).
}
\label{diff}
\begin{tabular}[c]{cccc}
\hline
\noalign{\smallskip}
$l$       &    $b$    & $R$        & $A_\mathrm{V}/R$     \\
(deg) & (deg) & (pc)  & (mag per kpc)   \\
\hline
\noalign{\smallskip}
180.0000 & $-89.6667$ & 475 & 0.000 \\
180.0000 & $-89.6667$ & 425 & 0.000 \\
180.0000 & $-89.6667$ & 375 & 0.020 \\
180.0000 & $-89.6667$ & 325 & 0.000 \\
180.0000 & $-89.6667$ & 275 & 0.000 \\
180.0000 & $-89.6667$ & 225 & 0.000 \\
180.0000 & $-89.6667$ & 175 & 0.000\\
180.0000 & $-89.6667$ & 125 & 1.000 \\
180.0000 & $-89.6667$ & 75  & 0.000 \\
180.0000 & $-89.6667$ &  25 & 1.560 \\
\ldots   &   \ldots   & \ldots & \ldots \\
\hline
\end{tabular}
\end{table*}


The AKQ22 data allow the spatial variations of the extinction law to be analyzed. For this purpose, we used the estimates of $A_\mathrm{V}$ in the $V$ filter and $A_\mathrm{G}$
in the Gaia $G$ filter from AKQ22 by modifying the latter estimates. For the $A_\mathrm{V}$ estimates AKQ22 and, subsequently, we adopted the extinction law from Schlafly et al. (2016), 
while for the modification of the $A_\mathrm{G}$ estimates we used the influence of the variations in the real extinction law on the effective temperature estimates obtained by AKQ22, 
as noted in their Section ``D.3. Variations in the extinction law induce systematic effective temperature shifts''. We used the fact that the main-sequence dwarfs being considered
by us exhibit a reliably determined average effective temperature--dereddened color relation. An analysis and allowance for this relation based on the AKQ22 data allowed us to 
calculate $A_\mathrm{G}$ as the reddening $E(G-W2)$ under the assumption of a negligible extinction in the WISE $W2$ IR filter with an effective wavelength of 4.6 microns. 
Thus, the modified $A_\mathrm{G}$ estimates are free from the assumption about the extinction law. Therefore, the spatial $A_\mathrm{G}/A_\mathrm{V}$ variations must reflect the 
real spatial variations of the extinction law. Note that we did not correct the $A_\mathrm{V}$ estimates for the systematics from Eqs. (1) and (2) by assuming the same systematics 
in the $A_\mathrm{G}$ estimates and, consequently, a negligible influence of this systematics on the ratio $A_\mathrm{G}/A_\mathrm{V}$.

However, this approach yields reliable results only in spatial cells with a large number of stars. Therefore, our map of $A_\mathrm{G}/A_\mathrm{V}$ variations has a lower relative
accuracy and a poorer actual resolution (though the formal resolution is the same) and covers the smaller space $R<800$ pc than does our 3D $A_\mathrm{V}$ map. An example of the 
$A_\mathrm{G}/A_\mathrm{V}$ variations with $l$ and $b$ within 800 pc of the Sun found by us is shown in Fig. 11. It is important that the $A_\mathrm{V}$ and $A_\mathrm{G}$ estimates 
under consideration differ not so much due to the difference between the $V$ and $G$ filters as due to the difference in allowance for the extinction law. Therefore, the derived
ratio $A_\mathrm{G}/A_\mathrm{V}$ can hardly be unambiguously converted to $R_\mathrm{V}$. However, the range $0.80<A_\mathrm{G}/A_\mathrm{V}<0.89$ roughly corresponds to the range 
$2<R_\mathrm{V}<4$.

Figure 11 shows significant variations of the extinction law with both $l$ and $b$. On the whole, the variations found agree with those found by Gontcharov (2012a, 2013, 2016a), 
including the regions of large $A_\mathrm{G}/A_\mathrm{V}$ at high latitudes. The vast region of reduced $A_\mathrm{G}/A_\mathrm{V}$ approximately between the points 
$l\approx130^{\circ}$, $b\approx+10^{\circ}$ and $l\approx200^{\circ}$, $b\approx-15^{\circ}$, which was previously found by Schlafly et al. (2016, 2017), is particularly
notable on our map. Thus, a more detailed fruitful analysis of the spatial variations in the extinction law based on the AKQ22 data and future Gaia data is possible in future.

\begin{figure*}
\includegraphics{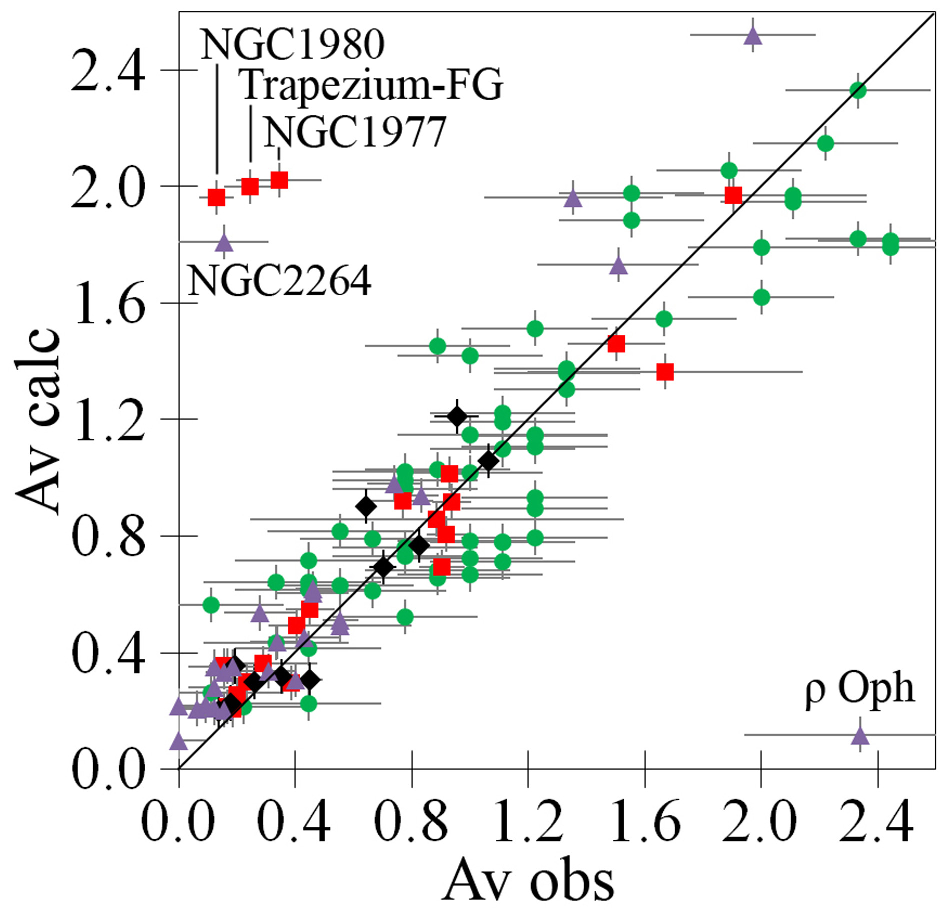}
\caption{Estimates of $A_\mathrm{V}$ for open clusters from the literature in comparison with the predictions of our 3D map: from Niu et al. (2020) - black diamonds, 
Monteiro et al. (2020) - red squares, He et al. (2021) - green circles, and Jackson et al. (2022) - violet triangles.
}
\label{openclusters}
\end{figure*}

\begin{figure*}
\includegraphics{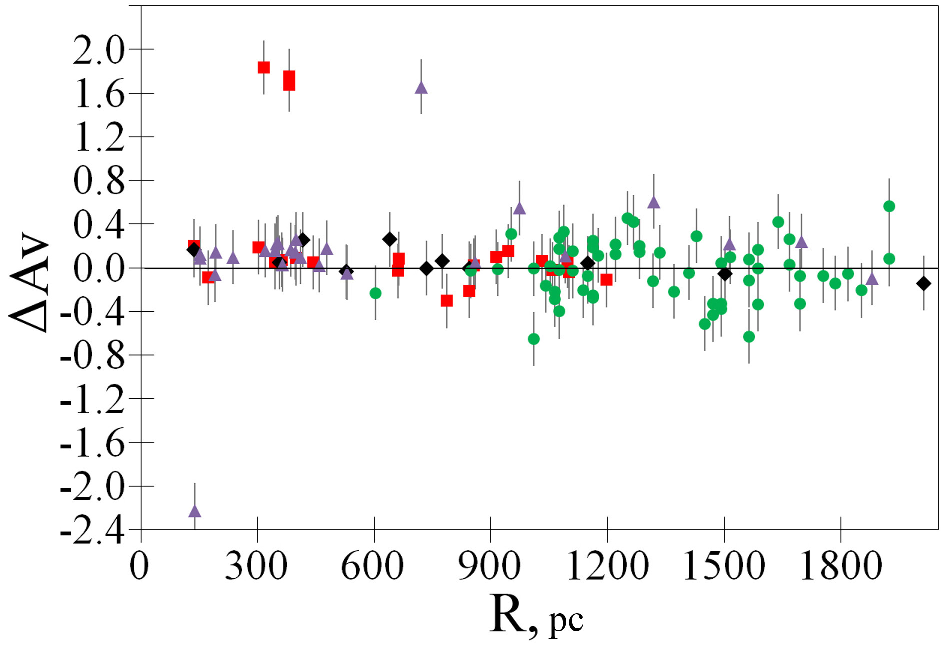}
\caption{Difference between the $A_\mathrm{V}$ estimates from our 3D map and the estimates from the literature for open clusters versus distance. 
The designations are the same as those in Fig. 12.
}
\label{rav}
\end{figure*}

\begin{figure*}
\includegraphics{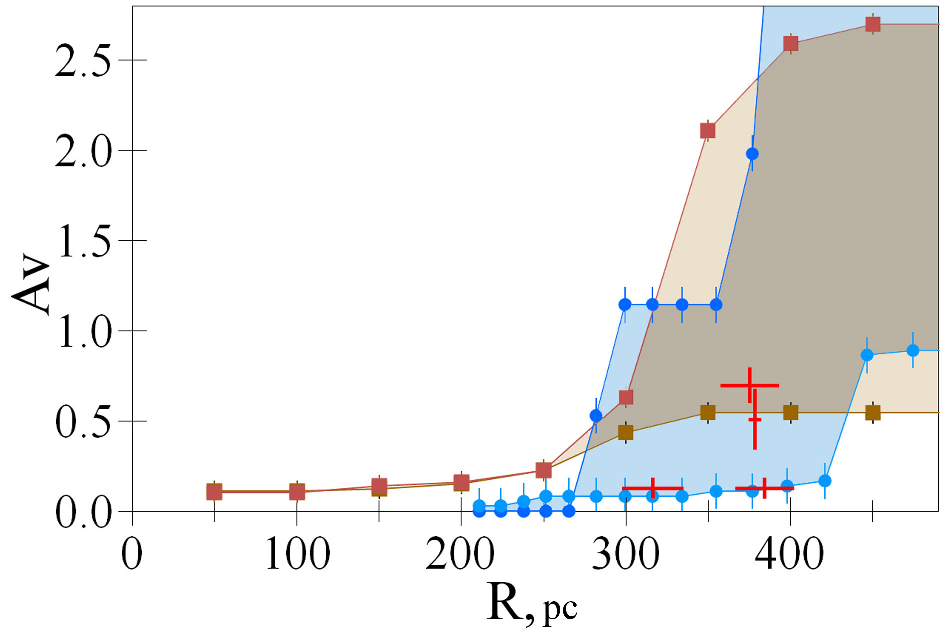}
\caption{The distance dependence of $A_\mathrm{V}$ in the NGC\,1980 field indicated by the brown squares of different shades for two lines of sight with the extreme estimates 
from our 3D map, the blue circles of different shades for two lines of sight with the extreme estimates from the GSZ19 map, and the red crosses whose size reflects the 
uncertainties declared by the authors for the independent estimates from Cantat-Gaudin et al. (2020), Hunt and Reffert (2023) and two estimates by different methods from 
Monteiro et al. (2020). The intermediate estimates from our map and GSZ19 on different lines of sight in the NGC\,1980 field fill the regions with the corresponding color.
}
\label{ngc1980}
\end{figure*}

\begin{figure*}
\includegraphics{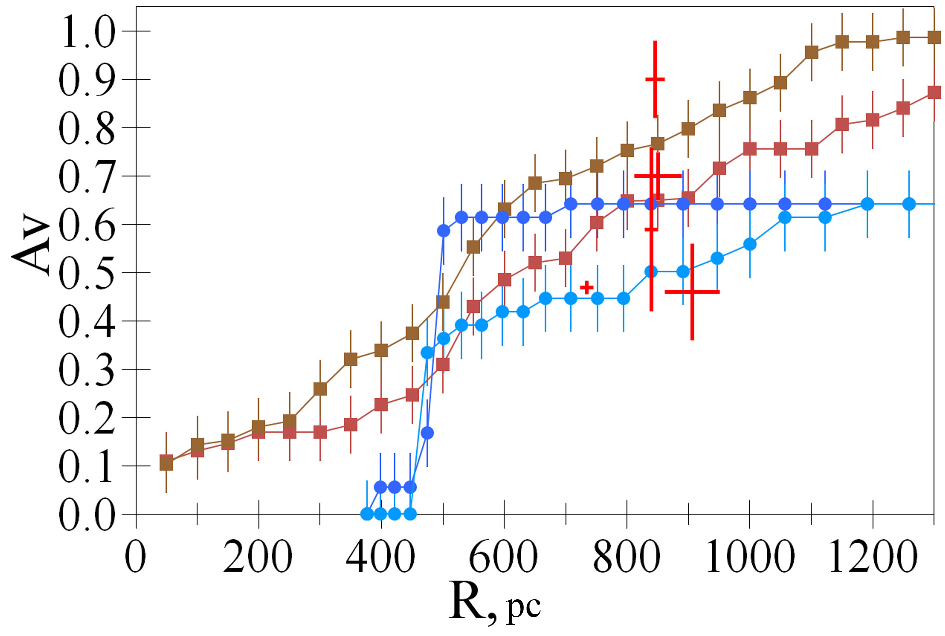}
\caption{The distance dependence of $A_\mathrm{V}$ in the NGC\,2168 field indicated by the brown squares of different shades for two lines of sight from our 3D map, the blue 
circles of different shades for two lines of sight from the GSZ19 map, and the red crosses whose size reflects the uncertainties declared by the authors for five independent 
estimates from Bossini et al. (2019), Cantat-Gaudin et al. (2020), Niu et al. (2020), Monteiro et al. (2020), and Hunt and Reffert (2023).
}
\label{ngc2168}
\end{figure*}

\section*{OPEN CLUSTERS}

The comparison of the predictions of our 2D map for globular clusters in the Section ``2D Map'' and Fig. 6 with the estimates from the literature yields unambiguous results, 
since the globular clusters are definitely outside the Galactic dust layer and have a negligible concentration of dust inside them. Using other test objects (open clusters, 
known dust clouds, variable stars, supergiants, etc.) requires a separate study, since the uncertainties in the characteristics of these objects (for example, the distances) 
are significant and make the comparison results ambiguous. To show the arising problems, we restricted ourselves to comparing our 3D map with some of the most accurate, in our view, 
extinction estimates for open clusters within 2 kpc of the Sun.

Figure 12 shows good agreement of the predictions of our model with the extinction estimates from Niu et al. (2020), Monteiro et al. (2020), He et al. (2021), and Jackson et al. (2022) 
obtained by different methods but without invoking any reddening/extinction maps and models. For their estimates He et al. (2021) adopted the uncertainty $\sigma(A_\mathrm{V})=0.25$; 
the uncertainties of the remaining estimates are provided by their authors. Just as for the globular clusters in Fig. 6, the open clusters in Fig. 12 show that our map may 
overestimate low extinctions and underestimate high one. However, this trend is better seen in Fig. 13, where the differences between our extinction estimates and those from 
the literature are shown as a function of distance. For distant clusters this trend may be explained by the selection in favor of the cluster members with a lower extinction, when
the members with a higher extinction are too faint to be observed.

Only five clusters show a great discrepancy between the predictions of the map and the estimates from the literature: NGC\,1977, NGC\,1980, and Orion Trapezium-FG 
(Monteiro et al. 2020), NGC\,2264, and $\rho$~Oph (Jackson et al. 2022). These clusters occupy a small part of the space inside large gas--dust clouds with regions of current 
star formation and large extinction gradients. The characteristics of these clusters are being actively refined, but are still known with a great uncertainty. For example,
owing to the detection of new members, the estimate of the NGC\,1980 radius has changed in recent years by almost a factor of 10: from 7.5 to 72 arcmin (8 pc) from the results 
of Bica et al. (2019) and Hunt and Reffert (2023), respectively. With the new estimate of the NGC\,1980 radius our maps give a set of extinction estimates in its field, and 
large extinction gradients become obvious. Figure 14 shows the dependence of $A_\mathrm{V}$ on distance $R$ for two lines of sight in the NGC\,1980 field with the extreme $A_\mathrm{V}$
estimates from our 3D map and the GSZ19 map. The various intermediate estimates on different lines of sights in the NGC\,1980 field, which are not shown for the sake of clarity, 
fill the regions with the corresponding color in Fig. 14. Independent estimates from the literature are shown for comparison. It can be seen from Fig. 14 that in the NGC\,1980 
field we can find the lines of sight on which the estimates of the maps agree at least with some estimate from the literature, while the great variety of estimates is probably 
caused by large extinction gradients in the NGC\,1980 field. This is true for all five mentioned clusters.

The extinction gradients in the fields of the remaining clusters are comparatively small, as shown, for example, by Fig. 15 for NGC\,2168. As a result, the estimates from our map 
and GSZ19 agree with the estimates from the literature no more poorly than those between themselves. It is possible that with such significant disagreement between the estimates
from the literature the estimates from our and other 3D reddening/extinction maps can be used to refine the distance and extinction estimates for some open clusters.

\section*{CONCLUSIONS}
\label{conclusions}

In this study we used the individual distance and interstellar extinction estimates for nearly 100 million dwarfs from AKQ22 based on Gaia DR3 parallaxes and Gaia, Pan-STARRS1, 
SkyMapper, 2MASS, and WISE photometry. As a result, we constructed five maps: (1) a 3D map of interstellar extinction $A_\mathrm{V}$ in the $V$ filter, (2) a 3D map of extinction 
$A_\mathrm{G}$ in the Gaia $G$ filter, (3) a 3D map of differential extinction $A_\mathrm{V}/R$ along line of sight -- all within 2 kpc of the Sun with a resolution of 50 pc and from
3.6 to 11.6 pc along and across the line of sight, respectively, (4) a 3D map of $A_\mathrm{G}/A_\mathrm{V}$ variations within 800 pc of the Sun, and (5) a 2D map of total Galactic
extinction $A_\mathrm{V}$ through the entire dust half-layer in the Galaxy from the Sun to extragalactic space with an angular resolution of 6.1 arcmin for Galactic latitudes
$|b|>13^{\circ}$.

In the AKQ22 estimates we found and took into account the systematic error in the extinction as a function of distance. Nevertheless, the AKQ22 estimates are probably the most 
accurate mass estimates of the individual extinctions for stars based on Gaia and up-to-date sky surveys. Therefore, in our study the AKQ22 data are particularly important as the
prototype of future Gaia results.

When producing the maps, we paid special attention to the space within 200 pc of the Sun and high Galactic latitudes as regions where the extinction estimates so far have had a 
large relative uncertainty. Our maps estimate the extinction within the Galactic dust layer from the Sun to an extended object or through the entire dust half-layer from the
Sun to extragalactic space with a precision $\sigma(A_\mathrm{V})=0.06$. This gives a high relative accuracy of extinction estimates even at high Galactic latitudes, where, 
according to our estimates, the median total Galactic extinction from the Sun to extragalactic objects is $A_\mathrm{V}=0.12\pm0.06$ mag. The accuracy of the predictions of our 
maps for a point object depends on the natural dust medium fluctuations and ranges from $\sigma(A_\mathrm{V})=0.08$ near the Galactic poles to $\sigma(A_\mathrm{V})=0.34$ 
near the Galactic equator. We showed that the presented maps are among the best ones in data amount, space size, resolution, accuracy, and other properties.

Our maps can be used to calibrate the 2DIR emission maps, to estimate the densities of large dust clouds and their distances (examples are given in Fig. 7), and to refine the 
characteristics of star clusters and other objects. Our 2D map is useful as a source of total Galactic extinction estimates for circum- and extragalactic objects, including globular
clusters, galaxies, quasars, and type Ia supernovae. Conversely, objects with independent accurate extinction and distance estimates can be used as tests to check the accuracy 
of the predictions of our maps.

\section*{ACKNOWLEDGMENTS}

We thank the referees for their useful remarks.

\section*{FUNDING}

This study was supported by the Russian Science Foundation (project no. 20-72-10052). 

The resources of the Centre de Donn\'ees astronomiques de Strasbourg, Strasbourg, France (http://cds.u-strasbg.fr), including the SIMBAD database and the X-Match service, 
were widely used in this study.
This research makes use of Filtergraph (Burger et al., 2013, an online data visualization tool developed at Vanderbilt University through 
the Vanderbilt Initiative in Data-intensive Astrophysics (VIDA) and the Frist Center for Autism and Innovation (FCAI, https://filtergraph.com).
This work has made use of data from the European Space Agency (ESA) mission Gaia (https://www.cosmos.esa.int/gaia), 
processed by the Gaia Data Processing and Analysis Consortium (DPAC, https://www.cosmos.esa.int/web/gaia/dpac/consortium).
This publication makes use of data products from the Wide-field Infrared Survey Explorer, which is a joint project of the University of California, 
Los Angeles, and the Jet Propulsion Laboratory/California Institute of Technology.
This study makes use of data products from the SkyMapper Southern Sky Survey. 
SkyMapper is owned and operated by The Australian National University's Research School of Astronomy and Astrophysics. 
The SkyMapper survey data were processed and provided by the SkyMapper Team at ANU. 
The SkyMapper node of the All-Sky Virtual Observatory (ASVO) is hosted at the National Computational Infrastructure (NCI).
This publication makes use of data products from the Two Micron All Sky Survey, which is a joint project of the University of Massachusetts and the Infrared 
Processing and Analysis Center/California Institute of Technology, funded by the National Aeronautics and Space Administration and the National Science Foundation.
This publication makes use of data products from the Pan-STARRS Surveys (PS1).

\section*{CONFLICT OF INTEREST}

The authors of this work declare that they have no conflicts of interest.

\section*{REFERENCES}


\newpage


\begin{thebibliography}{99}

\bibitem{amores2005} E.~B.~Am\^ores and J.~R.~D.~L\'epine, Astron. J. \textbf{130}, 659 (2005).

\bibitem{anders2022} F.~Anders, A.~Khalatyan, A.~B.~A.~Queiroz, C.~Chiappini, J.~Ard\`evol, L.~Casamiquela, F.~Figueras, \'O.~Jim\'enez-Arranz, C.~Jordi et al.,  
Astron. Astrophys. \textbf{658}, A91 (2022).

\bibitem{anderson2008} J.~Anderson, A.~Sarajedini, L.~R.~Bedin, I.~R.~King, G.~Piotto, I.~N.~Reid, M.~Siegel, S.~R.~Majewski, N.~E.~Q.~Paust et al., Astron. J. \textbf{135}, 2055 (2008).

\bibitem{baumgardt2021} H.~Baumgardt and E.~Vasiliev, MNRAS \textbf{505}, 5957 (2021)

\bibitem{bellazzini2002} M.~Bellazzini, F.~R.~Ferraro, and R.~Ibata, Astron. J. \textbf{124}, 915 (2002).

\bibitem{berry2012} M.~Berry, Z.~Ivezi\'c, B.~Sesar, M.~Juri\'c, E.~F.~Schlafly, J.~Bellovary, D.~Finkbeiner, D.~Vrbanec, T.~C.~Beers et al., Astrophys. J. \textbf{757}, 166 (2012).

\bibitem{bica2019} E.~Bica, D.~B.~Pavani, C.~J.~Bonatto, and E.~F.~Lima, Astron. J. \textbf{157}, 12 (2019).

\bibitem{bonatto2013} C.~Bonatto, F.~Campos, and S.~O.~Kepler, MNRAS \textbf{435}, 263 (2013).

\bibitem{bossini2019} D.~Bossini, A.~Vallenari, A.~Bragaglia, T.~Cantat-Gaudin, R.~Sordo, L.~Balaguer-N\'u\~nez, C.~Jordi, A.~Moitinho, C.~Soubiran et al., Astron. Astrophys. \textbf{623}, A108 (2019).

\bibitem{bressan} A.~Bressan, P.~Marigo, L.~Girardi, B.~Salasnich, C.~Dal Cero, S.~Rubele, and A.~Nanni, MNRAS \textbf{427}, 127 (2012).

\bibitem{filtergraph} D.~Burger, K.~G.~Stassun, J.~Pepper, R.~J.~Siverd, M.~Paegert, N.~M.~De Lee, and W.~H.~Robinson, Astronomy and Computing \textbf{2}, 40 (2013).

\bibitem{cantat2020} T.~Cantat-Gaudin, F.~Anders, A.~Castro-Ginard, C.~Jordi, M.~Romero-G\'omez, C.~Soubiran, L.~Casamiquela, Y.~Tarricq, A.~Moitinho et al., Astron. Astrophys. \textbf{640}, A1, (2020).

\bibitem{ccm89} J.~A.~Cardelli, G.~C.~Clayton, and J.~S.~Mathis, Astrophys. J. \textbf{345}, 245 (1989).

\bibitem{casagrande2014} L.~Casagrande and D.~A.~VandenBerg, MNRAS \textbf{444}, 392 (2014).

\bibitem{chambers2016} K.~C.~Chambers, E.~A.~Magnier, N.~Metcalfe, H.~A.~Flewelling, M.~E.~Huber, C.~Z.~Waters, L.~Denneau, P.~W.~Draper, D.~Farrow et al., arXiv:1612.05560 (2016).

\bibitem{clementini2022} G.~Clementini, V.~Ripepi, A.~Garofalo, R.~Molinaro, T.~Muraveva, S.~Leccia, L.~Rimoldini, B.~Holl, G.~Jevardat~de~Fombelle et al., Astron. Astrophys. \textbf{674}, A18 (2023).

\bibitem{dame2001} T.~M.~Dame, Dap~Hartmann, and P.~Thaddeus, Astrophys. J. \textbf{547}, 792 (2001).

\bibitem{dotter2011} A.~Dotter, A.~Sarajedini, and J.~Anderson, Astrophys. J. \textbf{738}, 74 (2011).

\bibitem{drimmel} R.~Drimmel, A.~Cabrera-Lavers, and M.~L{\'o}pez-Corredoira, Astron. Astrophys. \textbf{409}, 205 (2003).

\bibitem{gaiaedr3} Gaia collaboration, Astron. Astrophys. \textbf{649}, A1 (2021a).

\bibitem{riello2021} Gaia collaboration, Astron. Astrophys. \textbf{649}, A3 (2021b).

\bibitem{lindegren2021} Gaia collaboration, Astron. Astrophys. \textbf{649}, A4 (2021c).

\bibitem{delchambre} Gaia collaboration, Astron. Astrophys. \textbf{674}, A31 (2023).

\bibitem{trilegal} L.~Girardi, M.~A.~T.~Groenewegen, E.~Hatziminaoglou, and L.~Da~Costa, Astron. Astrophys. \textbf{436}, 895, (2005).

\bibitem{rcg} G.~A.~Gontcharov, Astron. Lett. \textbf{34}, 785 (2008).

\bibitem{gould} G.~A.~Gontcharov, Astron. Lett. \textbf{35}, 780 (2009).

\bibitem{rgb} G.~A.~Gontcharov, Astron. Lett. \textbf{37}, 707 (2011).

\bibitem{rv} G.~A.~Gontcharov, Astron. Lett. \textbf{38}, 12 (2012a).

\bibitem{av} G.~A.~Gontcharov, Astron. Lett. \textbf{38}, 87 (2012b).

\bibitem{ob} G.~A.~Gontcharov, Astron. Lett. \textbf{38}, 694 (2012c).

\bibitem{g2013} G.~A.~Gontcharov, Astron. Lett. \textbf{39}, 550 (2013). 

\bibitem{g2016} G.~A.~Gontcharov, Astron. Lett. \textbf{42}, 445 (2016a). 

\bibitem{astroph} G.~A.~Gontcharov, Astrophysics, \textbf{59}, 548 (2016b).

\bibitem{g17} G.~A.~Gontcharov, Astron. Lett. \textbf{43}, 472 (2017).

\bibitem{g19} G.~A.~Gontcharov, Astron. Lett. \textbf{45}, 605 (2019).

\bibitem{gm2017} G.~A.~Gontcharov and A.~V.~Mosenkov, MNRAS \textbf{470}, L97 (2017a).

\bibitem{gm2017big} G.~A.~Gontcharov and A.~V.~Mosenkov, MNRAS \textbf{472}, 3805 (2017b). 

\bibitem{gm2018} G.~A.~Gontcharov and A.~V.~Mosenkov, MNRAS \textbf{475}, 1121 (2018). 

\bibitem{polarization} G.~A.~Gontcharov and A.~V.~Mosenkov, MNRAS \textbf{483}, 299 (2019).

\bibitem{ngc5904} G.~A.~Gontcharov, A.~V.~Mosenkov, and M.~Yu.~Khovritchev, MNRAS \textbf{483}, 4949 (2019).

\bibitem{ngc6205} G.~A.~Gontcharov, M.~Yu.~Khovritchev, and A.~V.~Mosenkov, MNRAS \textbf{497}, 3674 (2020).

\bibitem{gm2021a} G.~A.~Gontcharov and A.~V.~Mosenkov, MNRAS \textbf{500}, 2590 (2021a). 

\bibitem{gm2021b} G.~A.~Gontcharov and A.~V.~Mosenkov, MNRAS \textbf{500}, 2607 (2021b). 

\bibitem{ngc288} G.~A.~Gontcharov, M.~Yu.~Khovritchev, A.~V.~Mosenkov, V.~B.~Il'in, A.~A.~Marchuk, S.~S.~Savchenko, A.~A.~Smirnov, P.~A.~Usachev, and D.~M.~Poliakov, MNRAS \textbf{508}, 
2688 (2021).

\bibitem{gms2022} G.~A.~Gontcharov, A.~V.~Mosenkov, S.~S.~Savchenko, V.~B.~Il'in, A.~A.~Marchuk, A.~A.~Smirnov, P.~A.~Usachev, D.~M.~Polyakov and N.Hebdon, Astron. Lett. \textbf{48}, 
578 (2022).

\bibitem{ngc6362} G.~A.~Gontcharov, M.~Yu.~Khovritchev, A.~V.~Mosenkov, V.~B.~Il'in, A.~A.~Marchuk, D.~M.~Poliakov, O.~S.~Ryutina, S.~S.~Savchenko, A.~A.~Smirnov et al., MNRAS 
\textbf{518}, 3036 (2023a).

\bibitem{ngc6397} G.~A.~Gontcharov, O.~S.~Ryutina, S.~S.~Savchenko, A.~V.~Mosenkov, V.~B.~Il'in, M.~Yu.~Khovritchev, A.~A.~Marchuk, D.~M.~Poliakov, A.~A.~Smirnov et al., MNRAS 
\textbf{526}, 5628 (2023b).

\bibitem{green2015} G.~M.~Green, E.~F.~Schlafly, D.~P.~Finkbeiner, H.-W.~Rix, N.~Martin, W.~Burgett, P.~W.~Draper, H.~Flewelling, K.~Hodapp et al., Astrophys. J. \textbf{810}, 25 (2015).

\bibitem{green2019} G.~M.~Green, E.~F.~Schlafly, C.~Zucker, J.~S.Speagle, and D.~P.~Finkbeiner, Astrophys. J. \textbf{887}, 93 (2019).

\bibitem{guo2021} H.-L.~Guo, B.-Q.~Chen, H.-B.~Yuan, Y.~Huang, D.-Z~Liu, Y.~Yang, X.-Y.~Li, W.-X.~Sun, X.-W.~Liu, Astrophys. J. \textbf{906}, 47 (2021).

\bibitem{hamrick2021} P.~Hamrick, A.~Bansal, and K.~Tock, Journal of the American Association of Variable Star Observers \textbf{49}, 192 (2021).

\bibitem{harris} W.~E.~Harris, Astron. J. \textbf{112}, 1487 (1996).

\bibitem{he2021} Z.-H.~He, Y.~Xu, C.-J.~Hao, Z.-Y.~Wu, and J.-J.~Li, Research in Astronomy and Astrophysics \textbf{21}, (93).

\bibitem{hunt2023} E.~L.~Hunt and S.~Reffert, Astron. Astrophys. \textbf{673}, A114, (2023).

\bibitem{jackson2022} R.~J.~Jackson, R.~D.~Jeffries, N.~J.~Wright, S.~Randich, G.~Sacco, A.~Bragaglia, A.~Hourihane, E.~Tognelli, S.~Degl'Innocenti et al., MNRAS \textbf{509}, 1664 (2022).

\bibitem{koch2014} A.~Koch and A.~McWilliam, Astron. Astrophys. \textbf{565}, A23 (2014).

\bibitem{lallement2022} R.~Lallement, J.~L.~Vergely, C.~Babusiaux, N.~L.~J.~Cox, Astron. Astrophys. \textbf{661}, A147 (2022).

\bibitem{mf2015} A.~M.~Meisner and D.~P.~Finkbeiner, Astrophys. J. \textbf{798}, 88 (2015).

\bibitem{monteiro2020} H.~Monteiro, W.~S.~Dias, A.~Moitinho, T.~Cantat-Gaudin, J.~R.~D.~L\'epine, G.~Carraro, and E.~Paunzen, MNRAS \textbf{499}, 1874 (2020).

\bibitem{niu2020} H.~Niu, J.~Wang, and J.~Fu, Astrophys. J. \textbf{903}, 93 (2020).

\bibitem{smss} C.~A.~Onken, C.~Wolf, M.~S.~Bessell, S.-W.~Chang, G.~S.~Da~Costa, L.~C.~Luvaul, D.~Mackey, B.~P.~Schmidt, and L.~Shao, Publications of the Astronomical Society of Australia \textbf{36}, 33 (2019).

\bibitem{panopoulou2022} G.~V.~Panopoulou, S.~E.~Clark, A.~Hacar, F.~Heitsch, J.~Kainulainen, E.~Ntormousi, D.~Seifried, and R.~J.~Smith, Astron. Astrophys. \textbf{663}, C1, (2022).

\bibitem{queiroz2018} A.~B.~A.~Queiroz, F.~Anders, B.~X.~Santiago, C.~Chiappini, M.~Steinmetz, M.~Dal~Ponte, K.~G.~Stassun, L.~N.~da~Costa, M.~A.~G.~Maia, et al., MNRAS \textbf{476}, 2556 (2018).

\bibitem{recio2005} A.~Recio-Blanco, G.~Piotto, F.~De~Angeli, S.~Cassisi, M.~Riello, M.~Salaris, A.~Pietrinferni, M.~Zoccali, A.~Aparicio, Astron. Astrophys. \textbf{432}, 851, (2005).

\bibitem{schlafly2011} E.~F.~Schlafly and D.~P.~Finkbeiner, Astrophys. J. \textbf{737}, 103 (2011).

\bibitem{schlafly2016} E.~F.~Schlafly, A.~M.~Meisner, A.~M.~Stutz, J.~Kainulainen, J.~E.~G.~Peek, K.~Tchernyshyov, H.-W.~Rix, D.~P.~Finkbeiner, K.~R.~Covey et al., Astrophys. J. \textbf{821}, 78 (2016).

\bibitem{schlafly2017} E.~F.~Schlafly,J.~E.~G.~Peek, D.~P.~Finkbeiner, G.~M.~Green, Astrophys. J. \textbf{838}, 36 (2017).

\bibitem{sfd98} D.~J.~Schlegel, D.~P.~Finkbeiner, and M.~Davis, Astrophys. J. \textbf{500}, 525 (1998).

\bibitem{2mass} M.~F.~Skrutskie, R.~M.~Cutri, R.~Stiening, M.~D.~Weinberg, S.~Schneider, J.~M.~Carpenter, C.~Beichman, R.~Capps, T.~Chester et al., Astron. J. \textbf{131}, 1163 (2006).

\bibitem{wagner2016} R.~Wagner-Kaiser, D.~C.~Stenning, A.~Sarajedini, T.~von~Hippel, D.~A.~van~Dyk, E.~Robinson, N.~Stein, and W.~H.~Jefferys, MNRAS \textbf{463}, 3768 (2016).

\bibitem{wagner2017} R.~Wagner-Kaiser, A.~Sarajedini, T.~von~Hippel, D.~C.~Stenning, D.~A.~van~Dyk, E.~Jeffery, E.~Robinson, N.~Stein, J.~Anderson, and W.~H.~Jefferys, MNRAS \textbf{468}, 1038 (2017).

\bibitem{wise} Wright~E.~L. et al., Astrophys. J. \textbf{140}, 1868 (2010).


\end{thebibliography}
\end{document}